\documentclass[final]{elsart}

\usepackage{amssymb}
\newcommand{\var}{\mathrm{Var}}

\newcommand{\tr}{\mathrm{tr}}

\begin{document}

\begin{frontmatter}



\title{Complementarity and Uncertainty in Mach-Zehnder Interferometry and beyond}

\corauth[cor1]{Corresponding author.}
\author{Paul Busch\corauthref{cor1}}
\ead{pb516@york.ac.uk}

\address{Department of Mathematics, University of York, UK,\\
and Perimeter Institute for Theoretical Physics, Waterloo, Canada}



\author{Christopher Shilladay}

\address{Department of Mathematics, University of Hull,  UK}
\ead{crshill@bridlington.u-net.com}

\begin{abstract}
A coherent account of the connections and contrasts between the principles of
complementarity and uncertainty is developed starting from a survey of the
various formalizations of these principles. The conceptual analysis is
illustrated by means of a set of experimental schemes based on Mach-Zehnder
interferometry. In particular, path detection via entanglement with a probe
system and (quantitative) quantum erasure are exhibited to constitute instances
of joint unsharp measurements of complementary pairs of physical quantities,
path and interference observables. The analysis uses the representation of
observables as positive-operator-valued measures (POVMs). The reconciliation of
complementary experimental options in the sense of simultaneous unsharp
preparations and measurements is expressed in terms of uncertainty relations of
different kinds. The feature of complementarity, manifest in the present
examples in the mutual exclusivity of path detection and interference
observation, is recovered as a limit case from the appropriate uncertainty
relation. It is noted that the complementarity and uncertainty principles are
neither completely logically independent nor logical consequences of one
another. Since entanglement is an instance of the uncertainty of quantum
properties (of compound systems), it is moot to play out uncertainty and
entanglement against each other as possible mechanisms enforcing
complementarity.
\end{abstract}


\end{frontmatter}
\newpage
{\parskip 2pt \tableofcontents }

\section{Introduction}\label{intro}

Soon after the inception of quantum mechanics, the notions of complementarity
and uncertainty were introduced to highlight features of the new theory, unknown
to classical physics, which amounted to limitations of the preparation and measurement
of atomic systems.

Complementarity and uncertainty continue to attract the attention
of researchers, inspiring novel experimental tests and demonstrations.
One such experiment was proposed by Scully, Englert and Walther
(SEW) \cite{SEW1991}. These authors described an atom interferometer in which
entanglement is utilized to store information about the path of an atom:  we will
use the term ``path marking" to refer to this process of path information storage in
a probe system by way of establishing correlations between the atom and the 
probe. SEW's claim is that in their
scheme, the standard explanation for the loss of interference upon
path marking in terms of classical random momentum kicks, and hence
the uncertainty relation, is not applicable.

The fact that in the SEW experiment the interference pattern was wiped out not
by classical kicks, which supposedly could be associated with an indeterminate
momentum, but by entanglement led to the suggestion that the principle of
uncertainty is less significant than complementarity: ``The
principle of complementarity is manifest although the position-momentum
uncertainty relation plays no role" (\cite{SEW1991}, p.~111); and in reply to
a critique \cite{STCW1994}, SEW go further, stating that ``the
principle of complementarity is much deeper than the uncertainty relation
although it is frequently enforced by $\delta x\delta p\ge\hbar/2$" \cite{SEW1995}.

A path-marking experiment similar to that proposed by SEW was experimentally
realized in 1998 by D\"{u}rr, Nonn and Rempe \cite{DNR1998} and confirmed the
conclusion of SEW, that neither mechanical disturbance nor the position-momentum uncertainty
relation could explain the loss of interference. This,
in turn, caused a controversy and led to grossly misleading articles in popular science
journals, announcing ``An end to uncertainty - Wave goodbye to the
Uncertainty Principle, you don't need it anymore." (New Scientist
1999) \cite{Buc1999}). Numerous papers appeared with conflicting conclusions
as to whether some forms of  random disturbance or uncertainty relations could always be
invoked to explain the loss of interference when path marking was effected via entanglement,
or whether entanglement was the more fundamental mechanism for this form of complementarity
(e.g. \cite{Wiseman-etal,LuisS-S,ESW2000,Bjork1999,KM2000,DR2000,Luis2001}).

The main aim of this paper is to study the roles and relative significance of the
uncertainty principle (in an appropriate understanding), ``disturbance" in measurement,
and entanglement in the explanation of complementary quantum phenomena, such as the
mutual exclusivity of path marking and interference detection.

It seems to us that the continuing lack of consensus over these questions is
largely due to persistent differences of general outlook on the principles of
complementarity and uncertainty that are manifest in the development just
reviewed, and due to the fact that these differences are usually not reflected
upon. The two notions are often considered as logically related, and the
uncertainty relation is frequently presented as a quantitative expression of
complementarity. By contrast, the work of SEW raised controversy over the
question whether the uncertainty relation is at all relevant to
complementarity. In effect, the considerations of SEW and the conclusions drawn
by D\"urr \emph{et al.} from their experiment suggest that complementarity and
uncertainty refer to aspects of quantum physics that can be discriminated
experimentally.

We therefore find it necessary to develop a coherent account of the notions and
principles of complementarity and uncertainty before we illustrate the
manifestations of these principles  in
the analysis of  a set of simple interferometric experiments.

To this end, we will briefly analyze the existing contrasting views on
complementarity and uncertainty, and review the development of various
formalizations of these so-called principles (Section \ref{sec:cu-rev}).
Separating the broader ``ideological" issues from the technical aspects will
help to elucidate the interesting specific physical and conceptual questions
indicated above, which can then be effectively addressed in terms of suitably
defined notions of complementarity and uncertainty.

The analysis of the experimental illustrations of these concepts will be carried out using the
general descriptions of observables in terms of positive-operator-valued 
measures (POVM). Projection-valued measures (PVMs) are
special cases of POVM and are called sharp observables. POVMs that
are not PVMs are called unsharp observables. Some of the experimental schemes
presented here are instances of joint measurements of an unsharp path and an
unsharp interference observable. The measurement theoretic concepts and tools
required for our analysis will be briefly reviewed in Section \ref{sec:tools}.

In Section \ref{sec:MZ}, we will discuss a range of atom and Mach-Zehnder
interferometric experiments. We will briefly review the type of which-path and
erasure experiments proposed by SEW and carried out by D\"urr \emph{et al.}
(subsection \ref{sec:SEW}). This will prepare the definition and study of
analogous Mach-Zehnder interferometric setups in subsection
\ref{sec:MZ-setup}-\ref{sec:quant eras}. The following experiments are
presented within one common setting and analyzed in detail: use of the
interferometer for path detection; an interference detection setup;
introduction of path marking via entanglement with a probe system, which
precedes the interferometer; quantum erasure; and quantitative quantum erasure.

The language of POVMs proves useful in section \ref{sec:c-u} for the
analysis of the relationships between specific versions of complementarity and
uncertainty as they manifest themselves in the present group of experiments.

General conclusions are drawn in Section \ref{sec:conc}. In the presentation of
complementarity and uncertainty developed here, these features are seen to be
logically related within quantum mechanics, although they have their separate
identities and roles. In a nutshell, \emph{uncertainty} is a direct consequence of the linear
structure of the Hilbert space formalism in that every vector state can be expressed
as a linear combination of eigenvectors of some observable, whose values are therefore
uncertain (or, more precisely, indefinite) in that state. This fact may be considered as
the broadest formulation of the \emph{uncertainty principle}, which is then usually
expressed as a trade-off relation between pairs of noncommuting observables. The
\emph{complementarity principle} highlights an extreme form of that relation, stating that
there are pairs of quantities that are such that complete
determination of one quantity implies maximal uncertainty about the other (and vice
versa). Thus, the formulation of complementarity presupposes the notion of uncertainty,
but the principles of complementarity and uncertainty cannot be reduced to one another.

Similarly, it is pointed out that \emph{entanglement} is an instance of uncertainty. It is
therefore pointless to attempt to separate entanglement as a ``mechanism"
enforcing complementarity in addition to, and  independently of, the uncertainty principle.

We hope that this study helps not only to settle some long-standing foundational issues
of quantum mechanics, but also to promote the idea that the time has come to
introduce the concept and application of POVMs in the undergraduate teaching of quantum
mechanics. POVMs are already being taught as a standard tool in
courses on quantum mechanics, quantum computation and quantum
information in a small but growing number of institutions around the world. The
concepts of the  underlying modern quantum theory of measurements are slowly
being brought to the attention of teachers of undergraduate quantum physics.
For example, simple implementations of POVMs for quantum optical experiments
were presented by H.~Brandt in the American Journal of Physics
\cite{Brandt1999}. There are a number of books giving elementary demonstrations
of the use of POVMs in the analysis of conceptual problems and quantum
experiments, the first being the fine text monograph by the late Asher Peres
\cite{Peres}. We dedicate this paper to the memory of Asher Peres, friend and
mentor to one of us (PB), and mediator between quantum foundations and quantum
applications.


\section{The notions of complementarity and uncertainty}\label{sec:cu-rev}

\subsection{Bohr, Heisenberg, and the consequences}\label{sec:cu-bird}

The principles of complementarity and uncertainty were introduced by Niels Bohr
\cite{Bohr1928} and Werner Heisenberg \cite{Heis1927} some eighty years ago in
their efforts to develop an intuitive understanding of quantum physics. Their
concern was to explain the dramatic deviations of this new theory from
classical physics, which manifested themselves in wave-particle duality and the
impossibility of defining and observing sharp particle trajectories.

From the beginning there was disagreement over the
relative significance and merits of the notions of complementarity and
uncertainty \cite{Jammer}. Bohr considered complementarity as fundamental for
his interpretation of quantum mechanics, which was to be based on classical concepts
and intuitions, rather than mathematical formalism. He thought of the uncertainty relation (primarily)
as a formal, quantitative expression of complementarity. By contrast, Heisenberg seemed to
have had little use for the notion of complementarity; he sought to develop an intuitive
understanding of quantum mechanics that derived from the formalism itself. In his
reminiscences, Heisenberg summarized the key that led him to the discovery of the
uncertainty relation in the famous sentence that he ascribed to Einstein: ``it is the theory
which decides what can be observed" \cite{Heis1977}.
It made him realize that there was room in quantum
mechanics for simultaneous approximate values of position and momentum, and thus for
unsharply defined trajectories as they are observed in cloud chambers.

Bohr and Heisenberg eventually reached a compromise on their divergent views in
the terminology of what came to be known as the Copenhagen interpretation of quantum mechanics.
However, the fundamental differences in their philosophical outlooks on quantum mechanics
were never resolved, as is manifest from their writings and recorded interviews and discussed, for example, in the book of Jammer \cite{Jammer}. The appearance of
unity maintained nevertheless by the Copenhagen pioneers has made it difficult to obtain a coherent account of the so-called Copenhagen interpretation, and there is a vast body of 
\emph{historico-philosophical} literature on this subject with rather divergent conclusions.

Here we are concerned with the reception of the notion of complementarity by
the \emph{physics} community. A survey of the textbook literature exhibits that three
points of view have evolved concerning the relationship and interplay between
the \emph{principles} of complementarity and uncertainty. However, it seems to
us that a fairly unambiguous, systematic account of the possible formalizations
of the \emph{notion} of complementarity within quantum mechanics has been
obtained.  We briefly recall the three distinct interpretational stances on the
status of the principle of complementarity, and then review  the possible
rigorous formalizations of the notion of complementarity.

There is one group of  authors who reiterate the account that has become part
and parcel of the so-called Copenhagen interpretation:
According to this view, the uncertainty relation constitutes the quantitative expression of the
principle of complementarity (e.g., \cite{Pauli1933}), and quantum mechanics
has been said to be best understood as \emph{the} theory of complementarity
\cite{Bohr1928,Pauli1933,Schwinger,Lahti1983}. Within this group, some describe the
uncertainty principle as \emph{enforcing} complementarity, while others deny the
uncertainty relation the status of principle, playing down its significance in
favor of complementarity \cite{Schwinger}. (Asher Peres had his own,
characteristically humorous way of doing this: in his textbook \cite{Peres},
the only occurrence of the term ``uncertainty principle" is in the index, with
the page reference indicating this very index page. On the other hand,
the principle of complementarity is also described only briefly, while there are
numerous occurrences of variants of uncertainty relations.)

A second group of authors (including, e.g., Dirac, von Neumann, Feynman) avoids
carefully any mention of the term complementarity, apparently in accordance with a
widespread perception that Bohr's presentations of this concept had remained
rather obscure. These authors invoke the uncertainty principle as the reason
for the impossibility of making simultaneous path and interference observations
with arbitrary precision. This approach is consistent with Heisenberg's point
of view, according to which it is possible but not necessary to refer to
complementarity for an intuitive interpretation of quantum mechanics.

Finally, today there is a widespread sense that complementarity
and uncertainty are best regarded as consequences of quantum mechanics which
highlight characteristic features of that theory but need not be held up as independent principles.

We will reassess these positions after the following brief review of the evolution of
current formulations of the complementarity and uncertainty principles.

\subsection{The notion of complementarity}\label{sec:c-review}

\subsubsection{Complementarity in Bohr's writings}

Bohr introduced the term \emph{complementarity} in his 1927
lecture in Como: ``The very nature of the quantum theory thus forces us to
regard the space-time co-ordination and the claim of causality, the union of
which characterizes the classical theories, as complementary but exclusive
features of the description, symbolizing the idealization of observation and
definition respectively." (\cite{Bohr1928}, p.~580).

Complementarity was thus originally conceived as a relationship between pairs
of descriptions, or phenomena, which are mutually exclusive but nevertheless
both required for a complete account of the physical system under
consideration. Bohr considered complementarity as a ``rational generalization"
of the classical notion of a causal spacetime description of physical
phenomena. He argued that in quantum physics, causality (represented by
conservation laws and deterministic equations of motion) and spacetime
description fall apart, as a consequence of what he called the quantum
postulate, according to which every observation (i.e., measurement) induces an
uncontrollable, unavoidable, and non-negligible change of the phenomenon (i.e.,
the observed system). In fact, as Bohr put it, spacetime coordination requires
observation and hence uncontrollable state changes, whereas a causal account of
phenomena requires the possibility of defining a state and hence precludes
interaction, and thus measurement.

For Bohr, the definition of a state and its observation constituted idealizations which were
simultaneously applicable to any desired degree of accuracy in classical physics.
According to the quantum postulate, the finite magnitude of Planck's
quantum of action introduces a limitation to the simultaneous applicability of
these idealizations: the influence of interactions required for an observation
is no longer negligible for atomic systems where the characteristic quantities
of the dimension of an action are comparable in magnitude to Planck's constant.
According to Bohr, such measurement interactions leave the object and measuring
apparatus in a situation that does not allow an independent description of
either system: ``Now, the quantum postulate implies that any observation of
atomic phenomena will involve an interaction with the agency of observation
not to be neglected. Accordingly, an independent reality in the ordinary physical
sense can neither be ascribed to the phenomena nor to the agencies of observation"
\cite{Bohr1928}. We agree with D.~Howard \cite{Howard}
that this passage should be read as expressing the
entanglement of object and apparatus after the measurement interaction.

Following Bohr's own practice, it has become customary to interpret his
informal descriptions of complementarity in terms of position and momentum
observables as follows: these observables are complementary in that for a
deterministic description of the trajectory of a particle, the values of both
observables are required, but according to the quantum mechanical limitations
of preparing and measuring the values of these observables, sharp values cannot
be assigned to them simultaneously. This impossibility of defining sharp
trajectories then gives room for the existence and explanation of interference
phenomena. A lucid account of complementarity along these lines was given by
W.~Pauli as early as 1933 \cite{Pauli1933}. In his reply of 1935 to the famous
challenge of Einstein, Podolsky and Rosen \cite{EPR1935}, Bohr defines the
complementarity of position and momentum \cite{Bohr1935}  by reference to ``the
mutual exclusion of any two experimental procedures, permitting the unambiguous
definition of complementary physical quantities." Complementarity thus
describes, according to Bohr, the limited way in which classical concepts can
be applied in the description of quantum experiments. Such a limitation is
imposed by quantum mechanics (the quantum postulate), in accordance with the
failure of classical physics to explain atomic phenomena. The importance of
complementarity derives, for Bohr, from the necessity of expressing all
physical experience in classical physical language.

\subsubsection{Complementarity and wave-particle duality}

It should be noted that in contrast to the classical exclusivity of wave and
particle theories that gave rise to the quantum puzzle of wave-particle
duality, the concepts of position and momentum \emph{are} simultaneously
applicable in classical physics but \emph{not} (without qualification) in
quantum mechanics. This implies that wave-particle duality cannot simply be
considered as an instance of complementarity, as is often suggested. There are
pairs of phenomena (and of related observables) that the same type of system
(e.g., electrons) can display and that are associated with intuitive ideas
relating to either particles or waves. But there are also instances of particle
and wave behaviors which do manifest themselves in the same experimental setup,
such as the Compton scattering of a photon with an electron observed through a
$\gamma$-ray microscope, followed by the subsequent wave-like propagation of
the light through the microscope, which accounts for the finite optical
resolution. In this and similar examples, Bohr freely used wave and particle
aspects in the analysis of one and the same phenomenon. Another simple example
is the interference observation in a double slit experiment where the
interference is explained by wave superposition while at the same time the
pattern is built up by the successive recordings of the photons or electrons
through their inelastic collisions with the molecules in the photographic
plate. According to Bohr, the apparent contradiction that lies in the
wave-particle duality is resolved through the realization of the limited
simultaneous applicability of classical concepts that is complementarity.

A dissolution, within quantum theory, of the problem of wave-particle
duality was carried out somewhat differently and more formally by Heisenberg in his 1929
Chicago lectures \cite{Heis1930}. He identifies particle and wave pictures with
the quantized theories of particles and fields, and sees the consistency of the
two pictures established by the equivalence of the quantum field theory,
restricted to the subspace of $N$ quanta, with the quantum mechanics of $N$
particles. On this account, there is no need for a simultaneous application of
wave and particle ideas since both formalisms can be used to give equivalent
accounts of the experiments in question.

These different assessments of the significance of wave-particle duality and of its
resolution in quantum mechanics are one example of the persistent discrepancies
between Bohr and Heisenberg. A detailed analysis of Heisenberg's account of
wave-particle duality, with similar conclusions to ours, has recently been given by
Camilleri \cite{Cam2006}.

\subsubsection{Modern formulations of complementarity}

The  complementarity idea was thus gradually transformed into the notion of
complementarity of  pairs of physical quantities, with an emphasis on the
\emph{negative} aspect of \emph{mutual exclusiveness} of value assignments or
experimental setups (such as path or interference detection). As will become
evident from the discussion of the uncertainty principle below, the
\emph{positive} aspect of \emph{mutual completion}, or \emph{complementation},
that comprises the original meaning of the term complementarity, has been
delegated to the uncertainty principle.

The statement of mutual exclusiveness of experimental setups can be interpreted
as constituting  three distinct forms of complementarity; one that refers to
the possibilities of state preparation, and two that refer to the possibilities
of joint and sequential measurements, respectively. This distinction allows one
to take into account the freedom in placing the Heisenberg cut between
preparation (object system) and measurement (probe system, measuring
apparatus), as well as the fact that often the state-preparing effects of
measurements are utilized. \emph{Preparation complementarity} is the
impossibility of preparing states in which the two observables in question are
simultaneously assigned sharp values. \emph{Measurement complementarity} is the
impossibility of performing simultaneous sharp measurements of these
observables, or the impossibility of performing their measurements in sequence
without mutual disturbance. In these general forms, complementarity applies to
practically all pairs of noncommuting  quantities (exceptions are pairs with
some common eigenstates).

Complementarity is usually understood to be a more specific
relationship that singles out certain important pairs of
observables, including, but not restricted to, canonically
conjugate pairs such as position and momentum, or components of angular momentum
and their associated angles. A comprehensive discussion of definitions of
preparation and measurement complementarity together with examples of
complementary pairs of observables are given in the monograph \cite{BGL1995} (Sec. IV.2)
and the review \cite{BuschLahti1995}. We will restrict ourselves here to a
brief summary with sufficient detail for the purposes of our subsequent analyses.

\subsubsection{Preparation complementarity}

The most widely adoped form of \emph{preparation
complementarity} (e.g.,  \cite{Schwinger,Kraus1987,SEW1991,Luis2001})
is  probably the following one that  we will refer to as \emph{value complementarity}
(following \cite{BGL1995}):
two observables are value complementary if whenever one has a definite
value, the values of the other are maximally uncertain. A value of observable
$A$ is definite if it occurs with probability equal to one in a measurement of
$A$, and the values of observable $B$ are maximally uncertain if they occur
with equal probabilities in a measurement of $B$.

Formally, the value
complementarity of two observables $A,B$ with discrete nondegenerate spectra
and associated eigenbasis systems $\{\psi_k:k=1,\ldots,n\}$, $\{\phi_\ell:\ell=1,\ldots,n\}$
amounts to the statement
that any two eigenstates have constant overlap, that is, the numbers $|\langle\psi_k|
\phi_{\ell}\rangle|$ are independent of $k,\ell$.  Pairs of orthonormal basis systems with this
property are called \emph{mutually unbiased}. It is well known that any observable $A$ with
nondegenerate eigenstates $\psi_1,\ldots,\psi_n$ has at least one partner $B$ with which it
forms a value complementary pair;
in fact there are infinitely many such partners: first one can
define any $B$ with orthonormal system of eigenstates
$\phi_{\ell}=\frac 1{\sqrt{n}}\sum_ke^{2\pi i  k\ell/n}\psi_k$, $\ell=1,\ldots,n$.
It is easily verified that $A,B$ are value complementary. Note that the eigenvalues of the
observables do not enter the definition of value complementarity. Further value-complementary
partners $B_U$ of $A$ are obtained by taking $B_U:=UBU^*$, where $U$ is any unitary operator
that commutes with $A$.

One must note that the notion of value complementarity does not easily extend
to continuous quantities or those with unbounded spectra. For example, in order
to consider position and momentum as value complementary, one must use their
improper eigenstates; one could try to capture the idea of nearly sharp
position values by means of sequences of normalizable states whose position
distributions approach a Dirac distribution. In that sequence, the momentum
distributions become arbitrarily widely spread out, but it is not clear how to
express the idea that these momentum distributions approach a uniform
distribution, which does not exist on the basis of normalizable states.

Similarly, in the case of number and conjugate phase, it is true that for a
number eigenstate, the phase is distributed uniformly, but there are no states
with definite, sharp values of the phase, nor are there states with uniform
number distributions \cite{BuLahPel}. To maintain the idea of value
complementarity, one would have to allow it to become a nonsymmetric relation.

These limitations do not arise if one adopts a slightly weaker form of
preparation complementarity, known as probabilistic complementarity.
Observables $A$, $B$ are \emph{probabilistically complementary}  if and only if
for their spectral projections $E^A(X)$, $E^B(Y)$ associated with bounded
intervals $X,Y$ (such that the projections $E^A(X)$, $E^B(Y)$  are different
from the null operator $O$ and the unit operator ${I}$) it is true that
whenever the probability $\mathrm{tr}[\rho E^A(X)]=1$ then
$0\ne\mathrm{tr}[\rho E^B(Y)]\ne 1$, and vice versa. This is to say that if the
value of $A$ is definitely in interval $X$, the value of $B$ is never certain
to be in any interval $Y$ (or its complement). This notion applies without
difficulty to the position-momentum and number-phase pairs; but it also allows
any two spin components of a spin-$\frac 12$ system to be complementary.

Probabilistic complementarity is known to be equivalent to the statement that for all
bounded intervals $X,Y$ (such that $E^A(X),E^B(Y)\ne O,I$), the associated spectral projections of
$A$ and $B$ satisfy
\begin{eqnarray}\label{eqn:AB-complem}
E^A(X)\land E^B(Y)&=O,\\
 E^A(X)\land E^B(\mathbb{R}\setminus Y)&=O,\\
E^A(\mathrm{R}\setminus X)\land E^B(Y)&=O.
\end{eqnarray}
Here, $P\land Q$ denotes the intersection of the projections $P$ and $Q$, that is, the projection
onto the intersection of their ranges; $R\setminus X$ denotes the complement of the set $X$. The equation $P\land Q=O$ is equivalent to the statement that
$P$ and $Q$ have no common eigenvectors associated with their eigenvalue 1.

\subsubsection{Measurement complementarity}

\emph{Measurement complementarity} can be specified in a similar vein to refer
to a pair of observables $A,B$ for which a sharp measurement of one of them
makes any attempt at measuring the other one simultaneously or in immediate
succession completely obsolete. The first form of measurement complementarity,
the impossibility of joint measurements, is a special instance of von Neumann's
theorem \cite{vN1932}, according to which two observables are jointly
measurable if, and only if, they commute. In a more specific, stronger form,
the measurement complementarity of two observables $A,B$ can be expressed as an
exclusion relation for the quantum operations describing the state changes due
to the measurements of $A$ and $B$. This characterization was again found to be
equivalent to the relation (\ref{eqn:AB-complem}) (see
\cite[Sec.~IV.2.3]{BGL1995}), thus demonstrating the match, stipulated already
by Bohr, between the possibilities of preparation and the possibilities of
measurement.

Measurement complementarity in the case of sequential measurements will
be taken to mean that due to the effect of the $A$ measurement, the $B$
measurement will not recover any information whatsoever about $B$ in the
(input) state immediately prior to the $A$ measurement. In extreme cases it may happen
that, although the second setup is devised to measure observable $B$, the statistics
obtained in the presence of the $A$ measurement is independent of the input
state; the observable effectively measured in the second measurement provides no information
about the system state prior to the $A$ measurement.
This is again a special instance of a general theorem
in the quantum theory of measurement (discussed, e.g., in \cite[pp.~42-44]{BGL1995}),
according to which a sequence of a measurement of observable $A$
followed immediately by a measurement set up to measure observable $B$
constitutes a joint measurement of $A$ and some unsharp observable $B'$ (represented
by a POVM) that commutes
with $A$. If $A$ and $B$ are value complementary observables with mutually
unbiased eigenbases and if $A$ is measured first by a von Neumann measurement, then
the observable $B'$ actually measured by a subsequent $B$ measurement scheme
is in fact trivial, in that its statistics carries no information about the system state prior
to the $A$ measurement.

Measurement complementarity is thus seen to reflect the fact that in quantum mechanics, 
every nontrivial measurement must alter the system's state, at least in the case of 
some input states. There can thus be no information gain without ``disturbance" in quantum 
mechanics. As will be discussed in the next subsection, the related issue of a measurement of 
one variable disturbing the distributions of other, noncommuting variables has been highlighted 
by Heisenberg by means of his famous thought experiments illustrating the uncertainty relation. 
This feature of quantum mechanics---the fact that measurements necessarily
alter the system under investigation, hence the impossibility to ``see" the system in its undisturbed form---has impressed the scientifically interested public as being 
of such fundamental importance that it is now widely known under the name ``Heisenberg effect" 
or also ``observer effect" (as is quickly confirmed by an internet search). In fact, these terms are 
used to denote loosely analogous phenomena  in a variety of disciplines, ranging from sociology,  political science or market research over anthropology and population ecology to computer science
(where certain forms of programming errors are called Heisenbugs).

We will see value complementarity and the two versions of measurement
complementarity (for joint or sequential measurements) at work in
the discussion of the Mach-Zehnder interferometry experiments; if the path observable
is represented by $\sigma_z$, any of its associated interference observables, represented
by $\cos\xi\,\sigma_x+\sin\xi\,\sigma_y$, is a complementary partner.

\subsubsection{Complementarity principle}

The \emph{complementarity principle} is the statement that there are
pairs of observables which stand in the relationship of
complementarity. As we saw above, this is satisfied in quantum mechanics for
observables the eigenvector basis systems of which are mutually
unbiased. We conclude that the ``principle" of complementarity, as
formalized here, is a consequence of the quantum mechanical formalism. There
seems to be no need to speak of a complementarity \emph{principle}, unless one
sets out to use such a principle in a more general framework to deduce quantum
mechanics. Here we follow the common practice of speaking of the complementarity
principle as a description of a remarkable feature of quantum mechanics.

\subsection{The notion of uncertainty and the uncertainty principle}\label{u-review}

\subsubsection{Origin of the uncertainty principle}

The uncertainty principle was introduced in Heisenberg's seminal
paper of 1927 \cite{Heis1927}; although he did not
speak of a principle, he made it very clear that he considered the
uncertainty relation as fundamental for an intuitive understanding
of quantum mechanics. He saw it as an expression of the physical
content of the canonical commutation relations for conjugate pairs
of quantities, and considered that these algebraic relations should
form the basis for a derivation of the formalism of quantum mechanics.

Heisenberg introduced quantum mechanical uncertainty with the
following words: ``It is shown that canonically conjugate
quantities can be determined simultaneously only with a characteristic inaccuracy," and
 ``\ldots the more precisely the position is determined
the less precisely the momentum is known and conversely."

Often the uncertainty principle is reduced to the idea, referred
to above as the Heisenberg effect, that any measurement ``disturbs" the
system in question. In the case of the canonically conjugate position and momentum
observables of a particle, this ``disturbance" is often identified
with a momentum kick imparted on the particle during the measurement.
If taken as generally valid, the explanation of the uncertainty principle in terms of
momentum kicks is an incorrect conflation whose origin must be
seen in Heisenberg's discussion of a position measurement with a $\gamma$-ray
microscope. Heisenberg pointed out that the higher the accuracy of the position
determination was, the shorter the wavelength of the photon, and therefore
the larger the momentum exchange with the observed particle. In a ``note added"
at the end of his 1927 paper,
he credits Bohr for the correction that the magnitude of the momentum transfer
did not constitute a cause for the particle's momentum to become indeterminate.
Bohr's explanation of the necessary momentum uncertainty was based on the dual nature
of the light used in the observation: the particle aspect of the photon
accounts for the momentum exchange resulting from Compton scattering, while the
wave aspect gives rise to an uncertainty in the momentum inference due to the finite
aperture of the microscope. Thus one could say that according to Bohr it is the quantum
nature of the probe system utilized in a measurement that enforces the necessary
uncertainty trade-off.

\subsubsection{Three varieties of uncertainty relations}

A careful reading of Heisenberg's 1927 paper
and his 1930 Chicago lecture notes \cite{Heis1930} shows that he
has in fact distinguished three variants of uncertainty relations.
It is evident that Heisenberg considered measurements to produce
(approximate) eigenstates of the measured observable,
corresponding to the measured value. Thus he describes the outcome
of an attempted joint measurement of position and momentum in
terms of the standard deviations of position and momentum
observables in a Gaussian wave function centered at the measured
values. The position and momentum uncertainties in this
conditional final state are then taken to represent the inaccuracies of
the joint measurement.

Here Heisenberg brings together two versions of uncertainty
relations: the uncertainty relation
\begin{equation}
\Delta(Q,\psi)\cdot\Delta(P,\psi)\ge\frac\hbar 2
\end{equation}
for state preparations,
according to which separate measurements of position and momentum
in a (vector) state $\psi$ have distributions with widths (standard deviations) satisfying
this uncertainty relation; and a trade-off relation
\begin{equation}
\delta q\cdot\delta p\ge\frac \hbar 2
\end{equation}
for the inaccuracies of joint measurements of these noncommuting observables.

Heisenberg did not have at his disposal a precise quantum-mechanical notion 
of joint measurement of noncommuting
observables. Such a notion was developed only several decades
later, after  POVMs had been introduced and made available
to describe unsharp or approximate measurements.
Heisenberg does grapple with the notion of joint
unsharp measurement and comes close to a solution by considering
sequences of measurements. For example, he considers the
diffraction of a matter wave at a slit and shows that if the
particle's momentum was initially sharp, this precision in the
definition of momentum becomes degraded during the passage through
the slit which effects an approximate localization of the
particle. Considered as a sequence of a sharp momentum measurement
followed immediately by an approximate position measurement, the
outcome of the sharp momentum determination is thus seen to be
modified into an unsharp momentum determination, due to the
``disturbing" influence of  the
approximate position determination. The resulting inaccuracies in
the definitions of position and momentum are shown to satisfy an
uncertainty relation. Indeed, the variance $\Delta(P,\psi)$ of momentum
after passage through the slit can be taken to represent both the
accuracy $\delta p$ of the momentum determination in the combined
measurement scheme and a measure of the disturbance $Dp$ of the (initially
sharp) momentum through the unsharp position determination:
$\Delta(P,\psi)=\delta p=Dp$. Likewise, the position uncertainty $\Delta(Q,\psi)$
upon passage through the slit reflects the accuracy $\delta q$ of the position
measurement. We thus have:
\begin{equation}
\delta q\cdot Dp\ge\frac\hbar 2.
\end{equation}

This version of uncertainty relation constitutes an accuracy-disturbance trade-off relation
for sequences of measurements. It has been carefully discussed in the context of
interference experiments by Pauli in his 1933 review \cite{Pauli1933}.
In the form described here, the disturbance
of the distribution of an observable $B$ through a measurement of $A$
is measured in terms of the variance of $B$ in the state immediately after the
selective $A$ measurement operation, which is to be compared to the
(near) zero variance of $B$ in an initial (near) eigenstate. In this formulation, the
accuracy-disturbance relation follows from the preparation uncertainty relation.
In a more general approach, the disturbance of the distribution of $B$ during a
measurement of $A$ should be described by some measure of the difference
between the distributions of $B$ before and after the nonselective $A$ measurement,
and it should depend on the accuracy of the $A$ measurement. Interestingly,
rigorous and general formulations of such disturbance uncertainty relations have been
investigated only rather recently  (e.g., \cite{Kraus1987,Ozawa,Wer}). A review of rigorous
formulations of all three types of uncertainty relations for position and momentum is given
in \cite{BuHeiLah2006}.

\subsubsection{Uncertainty principle}

We propose that the term \emph{uncertainty principle} refers to
the broad statement that there are  pairs of observables for which
there is a trade-off relationship in the degrees of sharpness of the
preparation or measurement of their values, such that a simultaneous
or sequential determination of the values requires a nonzero amount of
unsharpness (latitude, inaccuracy, disturbance). This comprises the
above three versions of uncertainty relations. There are a variety of
measures of uncertainty, inaccuracy, and disturbance with which
such trade-off relations can be formulated, usually in the form of
inequalities.

The term ``principle" refers here to the fact that the uncertainty relations
highlight an important nonclassical feature of quantum mechanics. They
are a formal consequence of the noncommutativity of the observables
in question. Being inequalities, the uncertainty relations can hardly be
considered adequate as postulates from which to derive quantum mechanics; 
however, they have been used to rule out the field of real numbers in favor 
of the complex numbers as the underlying field for the Hilbert space formulation
of quantum mechanics \cite{LahtiMac}.

\subsubsection{Uncertainty or indeterminacy?}

A discussion of the uncertainty principle would not be complete without a
comment on the nature of quantum mechanical uncertainty. The fact that many
observables do not have definite values even in pure states, which represent
maximal available information about the physical system under consideration,
stands in strong contrast to the situation in classical physics. If in a pure
state $\psi$ the probability for a property represented by a projection $P$ is
neither one nor zero, the value of any physical quantity represented by that
projection must be considered to be \emph{objectively indeterminate}, or
\emph{indefinite}, rather than only \emph{subjectively unknown}, or
\emph{uncertain}. According to the Kochen-Specker theorem (discovered by Kochen and 
Specker \cite{KS} and independently by Bell \cite{Bell1966}), any attempt at
hypothetically assigning definite values to sets of non-commuting observables
leads to contradictions even in the case of rather small, finite sets of such
observables. (Asher Peres was one of the champions in finding smaller and
smaller sets of observables with Kochen-Specker contradictions; see his book
\cite{Peres}.) The assignment of definite values to the position and momentum
of a particle in the Bohm interpretation is no exception to this conclusion
since these value attributions are contextual, that is, they are different for
different measurement setups. The conclusion is that the term
\emph{indeterminacy principle} appears to be more appropriate. However,
``\emph{uncertainty} principle" has become the standard name and shall be used
here, with the proviso that one should beware the misleading connotation of
subjective ignorance that it carries.

\subsection{Complementarity vs. uncertainty?}\label{sec:c vs u}

The above review shows that the concepts of complementarity and uncertainty
highlight two aspects of one and the same feature of quantum mechanics:

\begin{enumerate}
\item[(I)\ ] the \emph{impossibility}
of assigning simultaneously sharp values to certain pairs of noncommuting observables, be it
by preparation or measurement;
\item[(P)] the \emph{possibility} of simultaneously assigning unsharp values to such observables
by preparation of measurement.
\end{enumerate}

One may say that in contrast to Bohr, who emphasized the \emph{negative} aspect
of complementarity in the sense of (I), Heisenberg moved further to make a
\emph{positive} statement in the form of uncertainty relations which, if
satisfied, enabled the option (P). We believe that this account is endorsed by
Bohr who occasionally refers to the uncertainty relation as expressing
complementarity, in the sense of (I), as well as (P). This seems evident from
the following passage in the published version of the Como lecture which we
quote in full length:
\begin{quote}
In the language of the relativity theory, the content of the relations (2) [the
uncertainty relations] may be summarized in the statement that according to the
quantum theory a general reciprocal relation exists between the maximum
sharpness of definition of the space-time and energy-momentum vectors
associated with the individuals. This circumstance may be regarded as a simple
symbolical expression for the complementary nature of the space-time
description and claims of causality. At the same time, however, the general
character of this relation makes it possible to a certain extent to reconcile
the conservation laws with the space-time co-ordination of observations, the
idea of a coincidence of well-defined events in a space-time point being
replaced by that of unsharply defined individuals within finite space-time
regions. (Bohr 1928 \cite[Sec.~2]{Bohr1928})
\end{quote}
For Bohr, the ``mutual completion" part of complementarity refers to the
necessity of making use of the exclusive descriptions (or observables) in
different experimental contexts that cannot be created simultaneously.
Incidentally, the above quote seems to constitute the first occurrence of the
term ``unsharp" in connection with the question of simultaneous preparation or
measurement of position and momentum, which today is formalized in terms of
unsharp observables, that is, POVMs.

It seems largely a matter of terminological or interpretational preference
which aspects of the two statements (I) and (P) are to be subsumed under the
complementarity principle or the uncertainty principle. However, the
formalizations of the features (I) and (P) reviewed in the preceding
subsections, which are those most commonly used in the recent research
literature, have clearly identified (I) as an expression of the idea of
complementarity and (P) as the essence of the uncertainty principle. This
constitutes a break with two older traditions which gave preference either to
the complementarity principle or the  uncertainty principle. It appears to us
that with this terminological shift, a more balanced assessment has been
achieved: compared to the view that emphasized complementarity over
uncertainty, the positive role of the uncertainty relations as enabling joint
determinations and joint measurements is now highlighted more prominently; and
even though it is true (as we show later) that the uncertainty statement
(P) entails (I) in a suitable limit sense, it is still appropriate to point out
the strict mutual exclusivity of sharp value assignments which, after all, is
the reason for the quest for an approximate reconciliation in the form of
simultaneous but unsharp value assignments.

Irrespective of the particular terminological or interpretational preference,
formalizing the respective statements (I) and (P) has opened up new and
interesting questions: (I) and (P)  have become claims that can or cannot be
proven as consequences of the theory, and it becomes possible to study the
logical relationships between these statements. Questions like these will be
studied in the remaining part of this paper with respect to Mach-Zehnder
interferometric or, more generally, qubit observables.

\section{Interlude: measurement theoretic concepts and tools}\label{sec:tools}

In this section we review the representation of observables as positive
operator valued measures (POVMs), their use in the definition of joint unsharp
measurement of noncommuting observables, and the measurement theoretic
implementation of a POVM. We restrict ourselves to observables with finitely
many values and some examples in the context of qubit observables, as they will
be used in the following sections.

\subsection{Effects and POVMs}

The standard representation of observables in quantum mechanics
uses self-adjoint operators, or equivalently, the associated spectral
measures, acting in the Hilbert space $\mathcal{H}$ associated with
the physical system under consideration.
For example, a standard observable $A$ with discrete
spectrum $\{a_1,a_2,\ldots,a_n\}$ has a spectral representation
of the form $A=\sum\limits_{k=1}^{n}a_k
P_k$. The eigenvalues $a_k$ and the associated mutually orthogonal
spectral projections $P_k$ define the spectral measure
\begin{equation}\label{eqn:PVM}
P:\,a_k\mapsto P_k,
\end{equation}
which satisfies the normalization condition
\begin{equation}\label{eqn:PV-n}
\sum_kP_k=I.
\end{equation}
A projection-valued map of the form (\ref{eqn:PVM}) with the property (\ref{eqn:PV-n})
is also  called a \emph{projection-valued measure} (PVM).
The probabilities for the outcomes (eigenvalues) $a_m$ in a
measurement of $A$ on state $\psi$ are given by
\begin{equation}
\langle\psi|P_m|\psi\rangle=:\langle P_m\rangle_{\psi}=\langle P_m\rangle.
\end{equation}
The normalization of probabilities for all unit vectors $\psi$ is ensured by the
normalization condition (\ref{eqn:PV-n}). Standard observable $A$ and PVM
$P:\,a_k\mapsto P_k$ will also be referred to as \emph{sharp observable}.

In generalizing the formalism  to include positive-operator-valued
measures (POVMs) as representations of imperfect or inaccurate
measurements, it is noted that probabilities can generally be
represented by expectation values of positive operators that are
not necessarily projection operators nor necessarily commuting.
Such operators are called effects. Hence, a
measurement with outcomes $\{\lambda_1,\lambda_2,\ldots,\lambda_m\}$ is
represented by a POVM
\begin{equation}\label{eqn:POVM}
E:\,\lambda_\ell\mapsto E_\ell,
\end{equation}
with a unique collection of effects $E_1,E_2,\ldots,E_m$. The normalization
condition
\begin{equation}\label{eqn:POV-n}
\sum_\ell E_\ell=I
\end{equation}
is required to ensure the normalization of the probability distributions
$\lambda_\ell\mapsto \langle\psi|E_\ell | \psi\rangle$ for each vector state $\psi$.
Often we will simply denote a POVM $E$ in terms of the set of its effects
$\{E_1,\ldots,E_m\}$.
A POVM which is not a PVM will be called  \emph{unsharp observable}.

A POVM $\lambda_\ell\mapsto E_\ell$ is taken to represent a coarse
grained, or smeared version of a sharp observable (\ref{eqn:PVM})
if there is a stochastic matrix $(w_{\ell k})$ ($w_{\ell k}\ge 0$,
$\sum_\ell w_{\ell k}=1$) such that
$E_\ell:=\sum_kw_{\ell k}P_k$. Such a POVM, which is always commutative, 
represents an approximate measurement of the sharp observable.

For example, the spectral representation of  the Pauli spin-1/2 operator
$\sigma_x$ in $\mathcal{H}=\mathbb{C}^2$ has the form,
\begin{equation}
\sigma_x=a_{1}P_{1}+a_{2}P_{2}
=\textstyle{\frac{1}{2}}(I+\sigma_x)-\frac{1}{2}(I-\sigma_x),
\end{equation}
with eigenvalues $a_1= 1,\ a_2=-1$ and spectral projections
$P_1=\frac{1}{2}(I+\sigma_x)$, $P_2=\frac{1}{2}(I-\sigma_x)$.

A smeared version of this spectral measure is obtained by applying a
stochastic matrix
\begin{equation}
(f_{jk}):={\frac{1}{2}}\left(\begin{array}{cc}1+f&1-f\\
1-f&1+f\end{array}\right)\qquad -1\leq f\leq 1.
\end{equation}
The smeared version of the spectral measure of $\sigma_x$ is
$F=\{F_1,F_2\}$ where
$F_{\ell}:=\sum\limits_{k=1}^{2}f_{\ell k}P_k$. This gives,
\begin{equation}
F_{1}=\textstyle{\frac{1}{2}}(I+f\sigma_x),\qquad
F_{2}=\textstyle{\frac{1}{2}}(I-f\sigma_x)
\end{equation}
Similarly, a smeared version of $\sigma_z$, $G=\{G_1,G_2\}$ can be
defined as,
\begin{equation}
G_{1}=\textstyle{\frac{1}{2}}(I+g\sigma_z),\qquad
G_{2}=\textstyle{\frac{1}{2}}(I-g\sigma_z)
\end{equation}
where $-1\leq g\leq 1$.

In this example, we see that the class of POVMs is wide enough to include
\emph{trivial} POVMs: these are composed of effects which are all multiples of
the unit operator $I$. The above POVM $F$ becomes trivial if we put $f=0$.
Trivial POVMs give probabilities that do not depend on the state; they provide
no information at all. We have already found it convenient to make reference to
observables represented as trivial POVMs in the preceding section.

\subsection{Joint measurability}

In a joint measurement of two observables $F$ and $G$, one sets out to infer
the values of these observables from the output readings. Thus for
every pair of values of $F$ and $G$ there has to be a pointer
value and the statistics of these pointer values should reproduce the
probabilities for the values of $F$ and $G$ in the object's input
state. Thus, there should be a POVM, $E$, whose probabilities
are the joint probabilities for the outcomes of $F$ and $G$.
This means that the probability distributions of F and G should be
obtained as marginal distributions of the probability distributions
of $E$.

Such a POVM, $E$, is called a \emph{joint observable} of $F$ and $G$;  observables
$F$ and $G$ are the \emph{marginals} of $E$.

According to a theorem of von Neumann \cite[Sec.~III.3]{vN1932} two sharp
observables have a (sharp) joint observable exactly when they commute. Thus two
non-commuting observables cannot be sharply measured together. However, it has
been found that smeared versions of two \emph{noncommuting} sharp observables
may have a joint observable. The pair  $F=\{\frac{1}{2}(I\pm f\sigma_z)\}$,
$G=\{\frac{1}{2}(I\pm g\sigma_z)\}$ are known \cite{BuShill2003} to have a
joint observable exactly when,
\begin{equation}\label{joint-mt-condit}
f^2+g^2\leq 1.
\end{equation}
Thus for two POVMs to be jointly measurable their degrees of unsharpness
$|f|,|g|$ must be limited by this trade-off inequality. In this case it is
straightforward to give an example of a joint observable $E$, assuming for
simplicity $0\leq f,g\leq 1$,
\begin{eqnarray}
E_{11}=\textstyle{\frac{1}{4}}(I+f\sigma_x+g\sigma_z)
\qquad E_{21}=\textstyle{\frac{1}{4}}(I-f\sigma_x+g\sigma_z)\\
E_{12}=\textstyle{\frac{1}{4}}(I+f\sigma_x-g\sigma_z) \qquad
E_{22}=\textstyle{\frac{1}{4}}(I-f\sigma_x-g\sigma_z).
\end{eqnarray}
 Each operator $E_{k\ell}$ is positive because the eigenvalues are
$\frac{1}{4}(1\pm|(f,g)|)=\frac{1}{4}(1\pm\sqrt{f^2+g^2})\ge 0$ due to
(\ref{joint-mt-condit}). Moreover, the marginality relations are fulfilled:
\begin{eqnarray}
E_{11}+E_{12}=F_1\qquad E_{21}+E_{22}=F_2\\
E_{11}+E_{21}=G_1\qquad E_{12}+E_{22}=G_2.
\end{eqnarray}
In sections \ref{sec:mark} to \ref{sec:quant eras} we will give
measurement implementations of similar joint observables.

\subsection{Measurement implementation of a POVM}

Next we give a brief explanation of how a POVM can be implemented
in a suitable measurement scheme.

In a measurement a probe is coupled with the object system by a
unitary evolution after which the probe pointer is read. In
addition, if the object system is still available after the
measurement interaction has ceased, one may also perform another
measurement on it.

The following situation is encountered in sections \ref{sec:mark} to
\ref{sec:quant eras}: a photon gets entangled with a probe system
and then passes through a Mach-Zehnder interferometer. Finally, a 
joint measurement is
made of a probe observable and a detector observable. The purpose
of this joint measurement is to obtain information about the
photon state immediately prior to  the interaction with the
probe and subsequent passage through the interferometer. Such
information is available in the form of the output probabilities
if these can be expressed in terms of the photon's input state.

Given that the initial state of the probe and  the interferometer settings are
fixed in each run, it follows that the output probabilities are
indeed the expectation values of a POVM for the photon input
state.

Let $\psi_i=\alpha|1\rangle+\beta|2\rangle$ ($|\alpha|^2+\beta|^2=1$)
denote the input state of the photon,
$|p_0\rangle$ the initial probe state, $U$ the unitary evolution
operator representing the probe and the passage through the
interferometer. Then the final state of the combined system
is $\Psi_f= U\psi_i\otimes|p_0\rangle$. On this the sharp output
observable with projections $M_{k\ell}=|k\rangle\langle
k|\otimes|r_{\ell}\rangle\langle r_{\ell}|$ is measured. Here
$|k\rangle$, $k=1,2$, are the eigenstates of an object
observable measured after the interaction with the probe, and
$|r_\ell\rangle$, $\ell=1,2$, are eigenstates of a
pointer or output observable of the probe. (In sections
\ref{sec:erasure} and \ref{sec:quant eras}, different choices
will be made for the  $|r_\ell\rangle$.) The
output probabilities are then
\begin{eqnarray}
\langle\Psi_f|M_{k\ell}|\Psi_f\rangle &=&\langle
p_0|\langle\psi_i|U^*M_{k\ell}U|\psi_i\rangle|p_0\rangle\cr
 &=&|\alpha|^2\langle
p_0|\langle1|U^*M_{k\ell}U|1\rangle|p_0\rangle +\alpha^*\beta\langle
p_0|\langle1|U^*M_{k\ell}U|2\rangle|p_0\rangle\quad\nonumber\\
&\ +&\beta^*\alpha\langle p_0|\langle2|U^*M_{k\ell}U|1\rangle|p_0\rangle
+|\beta|^2\langle
p_0|\langle2|U^*M_{k\ell}U|2\rangle|p_0\rangle.\qquad\nonumber\\
\end{eqnarray}
These can be written as
\begin{equation}\label{eqn:app1}
\langle\Psi_f|M_{k\ell}|\Psi_f\rangle=|\alpha|^2
E_{k\ell}^{11} +\alpha^*\beta E_{k\ell}^{12}+\beta^*\alpha
E_{k\ell}^{21}+|\beta|^2 E_{k\ell}^{22}.
\end{equation}
Being probabilities, these numbers are non-negative and hence
the expression (\ref{eqn:app1}) is a quadratic form for the
variables $\alpha$ and $\beta$. That is to say, for each
$k,\ell$, the matrix $(E_{k\ell}^{ij})_{i,j=1,2}$ is positive semi-definite;
thus. it represents a positive operator
$E_{k\ell}$ defined in the Hilbert space of the photon. Hence,
\begin{equation}\label{in-out}
\langle\Psi_f|M_{k\ell}|\Psi_f\rangle=\langle\psi_i|E_{k\ell}|\psi_i\rangle
\end{equation}
for all $\psi_i$. Normalization of the output probability entails
$\sum\limits_{k\ell}E_{k\ell}=1$. Thus $(k,\ell)\mapsto E_{k\ell}$ is a
POVM, which represents the (input) observable of the object system
measured by the measurement scheme using the output observable
$(k,\ell)\mapsto M_{k\ell}$. Formula (\ref{in-out}) is the basis for the analysis of all the
measurement schemes discussed in the coming sections.

The above consideration illustrates a general theorem that states that every
measurement scheme defines a POVM for the object system. The converse statement
is also true: every POVM admits an implementation in terms of a measurement scheme.
For a general introduction to these results and for original references, the reader
may wish to consult \cite{BLM1991}. To our knowledge, the first proposals of realistic
models of joint measurements of unsharp qubit observables were developed in
\cite{Busch1987}. Further detailed
examples of measurement implementations of POVMs are given, for instance, in the monograph
\cite{BGL1995} and in the book of de Muynck \cite{deMuynck-book}.

\section{Path marking and erasure in an atom and Mach-Zehnder interferometry}\label{sec:MZ}

\subsection{The atom-interferometric experiment of SEW}\label{sec:SEW}

In a two slit atom interferometer \cite{SEW1991} each atom is
prepared in a superposition
\begin{equation}\label{eqn:sew2.0}
\psi_0(\mathbf{r})=\textstyle{\frac{1}{\sqrt{2}}}[\psi_1(\mathbf{r})+\psi_2(\mathbf{r})]
\end{equation}
of path states $\psi_1(\mathbf{r})$ and $\psi_2(\mathbf{r})$ which
represent the passage through the two slits. On the far capture screen (position coordinate
$\mathbf{R}$) an interference pattern will be observed according to
\begin{equation}\label{eqn:sew2.1}
P_0(\mathbf{R})=|\psi_0(\mathbf{R})|^2=
\textstyle{\frac{1}{2}}[|\psi_1|^2+|\psi_2|^2+\psi^*_1\psi_2+\psi^*_2\psi_1].
\end{equation}
SEW consider the situation in which atoms, prepared in an excited internal
state $|a\rangle$ and propagating in a superposition of states corresponding to
two collimated beam paths, arrive singly at micro-maser cavities preceding each
of the double slits \cite[Fig.~3]{SEW1991}. Once in the cavity, the atoms will
make a transition $|a\rangle\to|b\rangle$, spontaneously emitting a microwave
photon. The state of atom plus field changes from
\begin{equation}\label{eqn:sew2.2}
\Psi_0(\mathbf{r})=\textstyle{\frac{1}{\sqrt{2}}}
\big[\psi_1(\mathbf{r})+\psi_2(\mathbf{r})\big]\,|a\rangle\,|0_10_2\rangle
\end{equation}
to the entangled state
\begin{equation}\label{eqn:sew2.3}
\Psi(\mathbf{r})=\textstyle{\frac{1}{\sqrt{2}}}\big[\psi_1(\mathbf{r})\,|b\rangle \,|1_10_2\rangle+
\psi_2(\mathbf{r})\, |b\rangle \,|0_11_2\rangle\big],
\end{equation}
where $|1_10_2\rangle$ and $|0_11_2\rangle$ represent the field states
corresponding to one photon in cavity 1 and none in cavity 2 and vice versa.
Thus, the micro-maser cavities act as \emph{which-way} detectors only if a photon left
in the cavity changes the electromagnetic field in a detectable way.

The probability density on the capture screen is given by
\begin{eqnarray}\label{eqn:sew2.4}
P(\mathbf{R})&=&\textstyle{\frac{1}{2}}[|\psi_1|^2+|\psi_2|^2
+\psi^*_1\psi_2\langle1_10_2|0_11_2\rangle
+\psi^*_2\psi_1\langle0_11_2|1_10_2\rangle]\nonumber\\
&=&\textstyle{\frac{1}{2}}[|\psi_1|^2+|\psi_2|^2].
\end{eqnarray}
The interference (cross) terms vanish because the field states
$|1_10_2\rangle$ and $|0_11_2\rangle$ are orthogonal.

SEW also consider the possibility of recovering coherence and
thus the interference pattern by deleting or \emph{``erasing"} the path
information left in the microwave cavity detectors \cite[Refs.~26-31]{SEW1991}.

To model this they consider a new arrangement whereby the two
cavities are separated by a shutter-detector combination
\cite[Fig.~5a]{SEW1991}. This allows for the radiation either
to be confined to the upper or lower cavity when the shutters are
closed or for the radiation to be absorbed by a detector behind
each shutter when it is opened. In the latter case the path
marking information can be said to be erased. This will be explained presently.

In the erasure experiment one makes use of the fact that the state
(\ref{eqn:sew2.3}) has the equivalent form
\begin{equation}\label{eqn:sew2.5}
\Psi(\mathbf{r})=\textstyle{\frac{1}{\sqrt{2}}}\big[\psi_+(\mathbf{r})\, |b\rangle\,|+\rangle
+\psi_-(\mathbf{r})\,|b\rangle \,|-\rangle\big],
\end{equation}
where
\begin{equation}
\psi_{\pm}=\textstyle{\frac{1}{\sqrt{2}}}[\psi_1(\mathbf{r})\pm\psi_2(\mathbf{r})],\quad
|\pm\rangle=\textstyle{\frac{1}{\sqrt{2}}}[|1_10_2\rangle\pm|0_11_2\rangle].
\end{equation}

To display the effects of the erasure, the experimental procedure
is described as follows: after an atom arrives on the far screen, the shutters
are opened and the state of the detector behind the
shutters, which may have changed from its initial state $|d\rangle$ to
a new state $|e\rangle$ orthogonal to $|d\rangle$, is
recorded. The possible transitions are as follows, reflecting the sensitivity of the
detector to the field state $|+\rangle$ rather than $|-\rangle$:
\begin{equation}
|+\rangle\,|d\rangle\to|00\rangle\,|e\rangle,\quad
|-\rangle\,|d\rangle\to|-\rangle\,|d\rangle.
\end{equation}

In half the observations the detector will be found in an excited
state indicating that there had been a photon in one of the
cavities which has been absorbed. In the remaining cases there is
no detection. Thus, the total state makes the following transition:
\begin{equation}\label{eqn:sew2.6}
\textstyle{\frac 1{\sqrt
2}}[\psi_+\,|b\rangle\,|+\rangle+\psi_-\,|b\rangle\,|-\rangle]|d\rangle
\rightarrow \textstyle{\frac 1{\sqrt
2}}\big[\psi_+\,|b\rangle\,|0_10_2\rangle\,|e\rangle+
 \psi_-\,|b\rangle\,|-\rangle\,|d\rangle\big].
\end{equation}
As seen in Eq.~(\ref{eqn:sew2.5}) the symmetric atom state
$\psi_+$ is coupled with the symmetric cavity field state; thus the state
of the atom arriving at the screen selected if the detector is found in state $|e\rangle$
is $\psi_+$. The probability density of those atoms will show the maxima and minima
(fringes) of an interference pattern,
$P_+(\mathbf{R})=|\psi_{+}(\mathbf{R})|^2=P_0(\mathbf{R})$ (Eq.~(\ref{eqn:sew2.1})).

Atoms arriving at the screen for which there is no corresponding
signal from the erasure detector  (i.e., the detector is found in $|d\rangle$)
will display ``anti-fringes", $P_-({\mathbf{R}})=|\psi_-(\mathbf{R})|^2$, corresponding
to the selected state $\psi_-$.

 If all the events are counted, irrespective of the erasure
 detector state, the distribution is,
\begin{equation}\label{eqn:sew2.7}
\textstyle{\frac{1}{2}P_+({\mathbf{R}})+\frac{1}{2}P_-({\mathbf{R}})}=
\textstyle{\frac{1}{2}}[|\psi_1|^2+|\psi_2|^2]=P(\mathbf{R}).
\end{equation}
The maxima of one pattern coincide with the minima of the other one,
washing out the fringes.

This consideration shows that in the entangled state, $\Psi(\mathbf{r})$
(Eq.~(\ref{eqn:sew2.3})), the information about path as well as
interference is fully available. Choosing to measure the path marking basis
states, $|1_10_2\rangle, |0_11_2\rangle$, of the probe yields which way
information. Measuring instead the field states $|+\rangle,|-\rangle$ allows
the recovery of interference fringes or anti-fringes, respectively. The two
options are mutually exclusive; in the first case it is the interference
information which is lost whereas in the second case it is path information
which is lost.

It should be noted that this situation is related to the Einstein,
Podolsky, Rosen (EPR) experiment (in Bohm's version for spin ${1}/{2}$
pairs) which also makes use of an entangled state with more than one
biorthogonal
decomposition as in Eqs. (\ref{eqn:sew2.3}) and (\ref{eqn:sew2.5}).

\subsection{Mach-Zehnder interferometer: basic setup}\label{sec:MZ-setup}

Consider a special case of the Mach-Zehnder interferometer in Fig.~1 with no
path marking and no phaseshifter. The two possible input states from I$_{1}$,
I$_{2}$ will be represented by orthogonal unit vectors $|1\rangle,|2\rangle$,
of a two dimensional Hilbert space, $\mathcal{H}=\mathbb{C}^2$. When a photon
entering via I$_1$ (represented by a ``path" state $|1\rangle$) arrives at the beam splitter
BS$_1$ its state is changed to an equally weighted superposition, with appropriate phase shift
by $\pi/2$ upon reflection; and similarly for an input state $|2\rangle$:
\begin{equation}
|1\rangle\to\textstyle{\frac 1{\sqrt2}}\big[|1\rangle+i|2\rangle\big],\quad
|2\rangle\to\textstyle{\frac 1{\sqrt2}}\big[i|1\rangle+|2\rangle\big].
\end{equation}

Arriving at detector D$_1$ will be a component via the path
I$_1$BS$_1$M$_1$ reflected by BS$_2$ carrying a total phase shift
of ${\pi}/2$ from M$_1$ plus ${\pi}/2$ from BS$_2$, and
a component via the path I$_1$BS$_1$M$_2$ transmitted by BS$_2$
also carrying a total phase shift of ${\pi}/2$ from BS$_1$
plus ${\pi}/2$ from M$_2$. Hence, detector D$_1$ will
register an output as these two components are in phase and interfere
constructively.

Arriving at detector D$_2$ will be a component via the path I$_1$BS$_1$M$_1$
transmitted by BS$_2$ carrying a total phase shift of ${\pi}/2$ from
M$_1$, and a component via the path I$_1$BS$_1$M$_2$ reflected by BS$_2$ carrying
a total phase shift of ${\pi}/2$ from BS$_1$ plus ${\pi}/2$ from
M$_2$ plus ${\pi}/2$ from BS$_2$. Hence, detector D$_2$ will register no
output as these two components are out of phase by $\pi$ and interfere destructively.

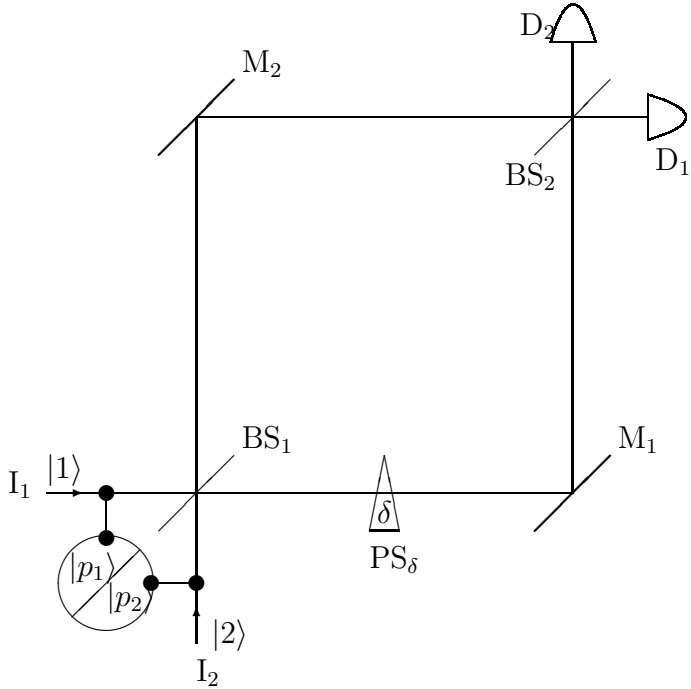
\begin{figure}
\setlength{\unitlength}{1.0mm}

\begin{picture}(120,90)(10,10)
\put(29,25){\line(1,1){10}} \put(79,75){\line(1,1){10}}
\put(14,30){\vector(1,0){5}\line(1,0){65}}
\put(34,80){\line(1,0){60}}
\put(34,10){\vector(0,1){5}\line(0,1){70}}
\put(84,30){\line(0,1){60}}
\put(57,25){\line(1,0){4}} \put(57,25){\line(1,5){2}}
\put(61,25){\line(-1,5){2}}
\qbezier(81,90)(84,100)(87,90)\put(81,90){\line(1,0){6}}
\qbezier(94,77)(104,80)(94,83)\put(94,77){\line(0,1){6}}
\put(22,18){\circle{12}} \put(22,18){\line(1,1){4.5}}
\put(22,18){\line(-1,-1){4.5}}
\put(22,24){\line(0,1){6}}\put(22,24){\circle*{2}}\put(22,30){\circle*{2}}
\put(28,18){\line(1,0){6}}\put(28,18){\circle*{2}}\put(34,18){\circle*{2}}
\put(9,30){I$_1$} \put(34,5){I$_2$} \put(14,32){$|1\rangle$}
\put(36,10){$|2\rangle$}
\put(40,36){BS$_1$} \put(75,71){BS$_2$} \put(40,86){M$_2$}
\put(90,36){M$_1$}
\put(58,26){$\delta$}
\put(17,19){$|p_1\rangle$} \put(22,15){$|p_2\rangle$}
\put(57,20){PS$_{\delta}$}
\put(95,73){D$_1$} \put(77,91){D$_2$}
\thicklines\put(29,75){\line(1,1){10}} \put(79,25){\line(1,1){10}}
\end{picture}
\vspace{8pt}

\caption{A Mach-Zehnder interferometer with path marking
and phase shifter.}\label{Fig1}

\end{figure}

So, if I$_1$BS$_1$M$_1$BS$_2$D$_1$ is the path represented by
$|1\rangle$, any measurement of the output of D$_1$ has associated
with it projector $|1\rangle\langle 1|$ representing one of the
eigenstates of the measured input observable according to Eq.~(\ref{in-out}) (which
defines the measured POVM). We identify this with the
spectral projection of the Pauli operator $\sigma_z$ associated
with the eigenvalue 1, $|1\rangle\langle1|=\frac{1}{2}(I+\sigma_z)$.

Similarly, I$_2$BS$_1$M$_2$BS$_2$D$_2$ corresponds to $|2\rangle$
and any measurement of the output of D$_2$ has associated with it the input
projector $|2\rangle\langle 2|=\frac{1}{2}(I-\sigma_z)$.

We are now in a position to consider a Mach-Zehnder interferometer with path marking
before the beam splitter BS$_1$. This will be implemented by
introducing a probe system which interacts with the photon. The
probe is represented by a two dimensional Hilbert space,
$\mathcal{H}=\mathbb{C}^2$, with path marking (``pointer") states $|p_1\rangle$
and $|p_2\rangle$ where $|p_1\rangle$ marks $|1\rangle$ and
$|p_2\rangle$ marks $|2\rangle$. A phase shifter, $\delta$ in one
path after BS$_1$ completes the analogy with the SEW experiment.

A general input from I$_{1}$, I$_{2}$ without the path marking interaction
switched on (object in input state $\psi_i$ and probe remaining in a neutral
state $|p_0\rangle$) can be represented by
\begin{equation}\label{mz eqn:3.1}
\psi_i\otimes|p_0\rangle
=(\alpha|1\rangle+\beta|2\rangle)\otimes|p_0\rangle
\end{equation}

Taking into account the phase shift in path 1 (see Fig.~\ref{Fig1}), the photon input
states $|1\rangle,|2\rangle$ undergo the following evolution upon passage through
the interferometer and before entering one of the detectors $D_1,D_2$:
\begin{eqnarray}\label{MZ-evol}
|1\rangle&\to \textstyle{\frac{1}{2}}[(-e^{i \delta}-1)
|1\rangle+i (e^{i \delta}-1)|2\rangle],\\
|2\rangle&\to \textstyle{\frac{1}{2}}[i (-e^{i \delta}+1)
|1\rangle-(1+e^{i \delta})|2\rangle].
\end{eqnarray}

When the path marking is turned on the total state after the path marking
interaction has ceased and before the photon enters BS$_1$ is
\begin{equation}\label{mz eqn:3.2}
\Psi_e=\alpha|1\rangle\otimes|p_1\rangle+\beta|2\rangle\otimes|p_2\rangle
\end{equation}
The photon states $|1\rangle,|2\rangle$ evolve according to (\ref{MZ-evol});
this leads to the total final (output) state
after the photon passes through beam splitter BS$_2$ as
\begin{eqnarray}\label{mz eqn:3.3}
\Psi_f^{\delta}&=\textstyle{\frac{1}{2}}\alpha[(-e^{i \delta}-1)
|1\rangle+i (e^{i \delta}-1)|2\rangle]\otimes|p_1\rangle\\
&\quad+\textstyle{\frac{1}{2}}\beta[i (-e^{i \delta}+1)
|1\rangle-(1+e^{i \delta})|2\rangle]\otimes|p_2\rangle\,.
\end{eqnarray}
We are now ready to discuss a variety of possible experiments.

\subsection{Path detection in outputs D$_{1}$,
D$_2$}\label{sec:none}

The simplest case of this Mach-Zehnder interferometer is with no path marking,
$|p_1\rangle=|p_2\rangle=|p_0\rangle$ and no phase shift,
$\delta=0$, analogous to a double slit interferometer (SEW) with
no path marking and the far field detector placed centrally; the
output state is,
\begin{equation}\label{mz eqn:3.4}
\Psi_f^o=-(\alpha|1\rangle+\beta|2\rangle)\otimes|p_0\rangle
\end{equation}
Observing the output of detectors D$_{1}$, D$_2$ with no path marking
is represented by the projections
$M_{1}=|1\rangle\langle1|\otimes{I}$,
$M_{2}=|2\rangle\langle2|\otimes{I}$.

The probabilities for an output at D$_1$ and D$_2$ are,
\begin{eqnarray}\label{mz eqn:3.5}
\langle\Psi_f^o|M_{1}|\Psi_f^o\rangle&=&\langle\psi_i|1\rangle\langle1|\psi_i\rangle
=|\alpha|^2\\
\langle\Psi_f^o|M_{2}|\Psi_f^o\rangle&=&\langle\psi_i|2\rangle\langle2|\psi_i\rangle
=|\beta|^2.
\end{eqnarray}
The input observable measured by this experiment is the POVM
$E^0=\{E^0_{1}$, $E^0_{2}\}$ defined by
\begin{equation}\label{mz eqn:3.6}
\langle\Psi_f^o|M_{k}|\Psi_f^o\rangle=\langle\psi_i|E_{k}^o|\psi_i\rangle
\end{equation}for all $\psi_i$ and $k=1,2$
It follows that $E^0$ is a PVM with projections
\begin{equation}\label{mz eqn:3.7}
E^0_1=\textstyle{\frac{1}{2}}(I+\sigma_z),\quad
E^0_2=\textstyle{\frac{1}{2}}(I-\sigma_z).
\end{equation}
This reproduces the discussion of path detection connected with Fig.~1:
If $\psi_i=|1\rangle$, then the probabilities of a detection at $D_1$ and $D_2$
are $\langle1|E_{1}^0|1\rangle=1$ and
$\langle1|E_{2}^0|1\rangle=0$, respectively. A similar consideration applies
to an input state $\psi_i=|2\rangle$. Thus, the measured observable is the path
observable $\sigma_z$.

\subsection{Interference detection at D$_1$, D$_2$}\label{sec:phase}

We now consider the use of the Mach-Zehnder interferometer for an
interference measurement. In a double slit
interferometer both slits would be open and a detector placed at
the first minimum. Here we choose $\delta=-{\pi}/2$; then the
output state is,
\begin{equation}\label{mz eqn:3.8}
\Psi_f^{{-{\pi}/2}}=\textstyle{\frac{-(1-i )}{\sqrt 2}}
\left[\alpha\textstyle{\frac{1}{\sqrt 2}}(|1\rangle-|2\rangle)+
\beta\textstyle{\frac{1}{\sqrt
2}}(|1\rangle+|2\rangle)\right]\otimes|p_0\rangle.
\end{equation}
If a measurement of $M_{k}=|k\rangle\langle k|\otimes I$,
$k=1,2$ is now applied by observing the outputs of D$_k$ the
associated probabilities are
\begin{eqnarray}\label{mz eqn:3.9}
\langle\Psi_f^{{-{\pi}/2}}|M_{1}|\Psi_f^{{-{\pi}/2}}\rangle
&=\textstyle{\frac{1}{2}}(|\alpha+\beta|^2)
=\langle\psi_i|E^{-{\pi}/2}_1|\psi_i\rangle\\
\langle\Psi_f^{{-{\pi}/2}}|M_{2}|\Psi_f^{{-{\pi}/2}}\rangle
&=\textstyle{\frac{1}{2}}(|\alpha-\beta|^2)
=\langle\psi_i|E^{-{\pi}/2}_2|\psi_i\rangle,
\end{eqnarray}
from which $E_{1}^{-{\pi}/2}$ and $E_{1}^{-{\pi}/2}$ are extracted:
\begin{equation}\label{mz-eqn:3.10}
E_{1}^{-{\pi}/2}=\textstyle{\frac{1}{2}(I+\sigma_x)},\quad
E_{2}^{-{\pi}/2}=\textstyle{\frac{1}{2}(I-\sigma_x)}.
\end{equation}
Following customary practice we consider an interference
observable one with the form
$\cos\delta\,\sigma_x+\sin\delta\,\sigma_y$, $0\le\delta<\pi$,
given that the path is represented by $\sigma_z$. Interference
observables are singled out by the condition that the interference
contrast can assume its maximum possible value. In this case their
eigenstates give equal probabilities of ${1}/2$ to the path
projections $|1\rangle\langle1|, |2\rangle\langle2|$.

In the present experiment, the measured input observable is defined by the
projections of Eq.~(\ref{mz-eqn:3.10}); these are the spectral projections of
the operator $\sigma_x$, which is indeed an interference observable.

\subsection{The path-marking setup}\label{sec:mark}

Now consider the case where $|p_1\rangle$ and $|p_2\rangle$ are
mutually orthogonal , $\langle p_1|p_2\rangle=0$. This is an
analog of SEW's path-marking experiment. We can find the
influence of the path marking on the outputs of the detectors
using each of the four measurement projections of path $|k\rangle$
jointly with marker $|p_\ell\rangle$, $M'_{k\ell}=|k\rangle\langle
k|\otimes |p_\ell\rangle\langle p_\ell |$, $k,\ell=1,2$, e.g.
\begin{equation}\label{mz eqn:3.11}
\langle\Psi_f^\delta|M'_{11}|\Psi_f^\delta\rangle=|\textstyle{\frac{1}{2}}
\alpha(e^{i \delta}+1)|^2=\textstyle{
\frac{1}{4}}|\alpha|^2(1+\cos\delta).
\end{equation}
The input POVM, the measured observable is again defined by Eq.~(\ref{in-out}),
\begin{equation}
\langle\Psi_f^\delta|M'_{k\ell}|\Psi_f^\delta\rangle=\langle\psi_i|E'_{k\ell}|\psi_i\rangle,
\end{equation}
and we obtain
\begin{eqnarray}\label{mz eqn:3.12}
E'_{11}&=\textstyle{\frac{1}{2}(I+\sigma_z)\cos^2\frac{\delta}{2}}\qquad
E'_{21}=\textstyle{\frac{1}{2}(I+\sigma_z)\sin^2\frac{\delta}{2}}\\
E'_{12}&=\textstyle{\frac{1}{2}(I-\sigma_z)\sin^2\frac{\delta}{2}}\qquad
E'_{22}=\textstyle{\frac{1}{2}(I-\sigma_z)\cos^2\frac{\delta}{2}}\,.
\end{eqnarray}
These effects are all fractions of a path projection.

We now give the marginal input POVM associated with the
detectors D$_{1}$, D$_{2}$,
\begin{eqnarray}\label{mz eqn:3.13}
F'_1&=E'_{11}+E'_{12}=\textstyle{\frac{1}{2}}(I+\cos\delta\,\sigma_z),\\
F'_2&=E'_{22}+E'_{21}=\textstyle{\frac{1}{2}}(I-\cos\delta\,\sigma_z).
\end{eqnarray}
This POVM represents a path observable smeared by $\cos\delta$. The
unsharpness inherent in the detector marginal is reflected in the non-zero
probability of the marker indicating path 1 but detector D$_2$ firing even if
the input state is a path eigenstate $|1\rangle$. Here we see the effect of the
perfect path-marking interaction: irrespective of the phase parameter value,
the Mach-Zehnder interferometer does not detect any interference effects.
When $\delta=0$, the POVM
$\{F'_{1},F'_2\}$ becomes a sharp path observable and when
the interferometer is set to observe interference, $\delta=\frac{\pi}{2}$, this POVM is
reduced to being trivial, $F'_{1}=\frac{1}{2}I=F'_2$, giving no path nor interference
information. This is in line with the prediction of SEW: path marking results
in the interference pattern being washed out. After the path marking
interaction, all the detectors are able to ``see" is a ``shadow" of the path
information provided by the path marker: indeed, the marginal POVM measured by
the path marker is given by the effects
\begin{eqnarray}\label{mz eqn:3.14}
G'_1&=E'_{11}+E'_{21}=\textstyle{\frac{1}{2}}(I+\sigma_z),\\
G'_2&=E'_{22}+E'_{12}=\textstyle{\frac{1}{2}}(I-\sigma_z).
\end{eqnarray}
These represent a sharp path observable irrespective of the value of $\delta$.

It is possible to define a third `marginal',
\begin{eqnarray}
H'_1&=E'_{11}+E'_{22}=\cos^2\textstyle{\frac{\delta}{2}}\,I,
\\
H'_2&=E'_{12}+E'_{21}=\sin^2\textstyle{\frac{\delta}{2}}\,I,
\end{eqnarray}
which in the present experiment also turns out to be trivial.

\subsection{Quantum erasure}\label{sec:erasure}

In the previous experiment, each path was correlated with one of
two orthogonal marker states. We can now consider a new set of
pointer states, which are
superpositions of the two orthogonal path-marker states,
\begin{equation}\label{mz eqn:3.15}
|q_1\rangle = \textstyle{\frac
1{\sqrt2}}(|p_1\rangle+e^{i \gamma}|p_2\rangle),\ \
|q_2\rangle=\textstyle{\frac 1{\sqrt
2}}(|p_1\rangle-e^{i \gamma}|p_2\rangle)\,.
\end{equation}
Observing these symmetric states involves outputs for which both
$|p_1\rangle$ and $|p_2\rangle$ are equally likely, so no information about the path
is recorded.

The final state (\ref{mz eqn:3.3}) in terms of the new pointer states is
\begin{eqnarray}\label{mz eqn:3.17}
\Psi_f^{\delta,\gamma}&=\textstyle{\frac 1{2\sqrt 2}}\left[\left(-\alpha(1+e^{i \delta})
+i  e^{-i \gamma}\beta(1-e^{i \delta})\right)|1\rangle\right.\nonumber \\
&\left. +\left(-i \alpha(1-e^{i \delta})-e^{-i \gamma}\beta(1
+e^{i \delta})\right)|2\rangle\right] \otimes|q_1\rangle\nonumber\\
& +\textstyle{\frac 1{2\sqrt 2}}\left[\left(-\alpha(1+e^{i \delta})-i  e^{
-i \gamma}\beta(1-e^{i \delta})\right)|1\rangle\right. \\
&\left.  +\left(-i \alpha(1-e^{i \delta})+e^{-i \gamma}\beta(1
+e^{i \delta})\right)|2\rangle\right] \otimes|q_2\rangle.\nonumber
\end{eqnarray}

As before we can find the four joint probabilities for the marker
and detector outputs, defined as the expectations of the
projections
 $M''_{k\ell}=|k\rangle\langle k|\otimes |q_\ell\rangle\langle q_\ell |$, $k,\ell=1,2$.

The associated input POVM $E''$ is determined via the relation
\begin{equation}
\langle\Psi_f^{\delta,\gamma}|M''_{k\ell}|\Psi_f^{\delta,\gamma}\rangle
=\langle\psi_i|E''_{k\ell}|\psi_i\rangle.
\end{equation}
We obtain:
\begin{eqnarray}\label{mz eqn:3.18}
E''_{11}&=&\textstyle{\frac{1}{4}}(I-\sin\delta\cos\gamma\,\sigma_x
-\sin\delta\sin\gamma\,\sigma_y+\cos\delta\,\sigma_z)
=\textstyle{\frac{1}{4}}(I-\mathbf{n}''\cdot\sigma)\, ,\\
E''_{21}&=&\textstyle{\frac{1}{4}}(I+\sin\delta\cos\gamma\,\sigma_x
+\sin\delta\sin\gamma\,\sigma_y-\cos\delta\,\sigma_z)
=\textstyle{\frac{1}{4}}(I+\mathbf{n}''\cdot\sigma)\, ,\\
E''_{12}&=&\textstyle{\frac{1}{4}}(I+\sin\delta\cos\gamma\,\sigma_x
+\sin\delta\sin\gamma\,\sigma_y+\cos\delta\,\sigma_z)
=\textstyle{\frac{1}{4}}(I+\mathbf{m}''\cdot\sigma)\, ,\\
E''_{22}&=&\textstyle{\frac{1}{4}}(I-\sin\delta\cos\gamma\,\sigma_x
-\sin\delta\sin\gamma\,\sigma_y-\cos\delta\,\sigma_z)
=\textstyle{\frac{1}{4}}(I-\mathbf{m}''\cdot\sigma)\, .
\end{eqnarray}

Here we have introduced the unit vectors
\begin{equation}\label{mz eqn:3.19}
\mathbf{n}''=(\sin\delta\cos\gamma,\sin\delta\sin\gamma,-\cos\delta),\
\mathbf{m}''=(\sin\delta\cos\gamma,\sin\delta\sin\gamma,\cos\delta).
\end{equation}

The marginal POVM associated with the detector outputs is obtained
by summing over both probe outputs:
\begin{equation}\label{mz eqn:3.20}
F''_{1}=E''_{11}+E''_{12}=\textstyle{\frac{1}{2}}(I+\cos\delta\,\sigma_z),\
F''_{2}=E''_{21}+E''_{22}=\textstyle{\frac{1}{2}}(I-\cos\delta\,\sigma_z).
\end{equation}

This is a smeared path observable. The marginal POVM associated
with the probe outputs is obtained by summing over both detection outputs:
\begin{equation}\label{mz eqn:3.21}
G''_{1}=E''_{11}+E''_{21}=\textstyle{\frac{1}{2}}\,I,\quad
G''_{2}=E''_{12}+E''_{22}=\textstyle{\frac{1}{2}}\,I.
\end{equation}
This is a trivial observable, it provides no information about the
input state $\psi_i$.

The fact that the detector POVM is a smeared path observable and
the probe POVM is trivial can be understood as follows. The
entanglement between probe and photon is devised to establish a
strict correlation between the path states $|1\rangle$,
$|2\rangle$ and the pointer states $|p_1\rangle$, $|p_2\rangle$,
for any photon input state $\psi_i$. This correlation information
is not accessible by measuring a probe observable with the
eigenstates $|q_1\rangle$, $|q_2\rangle$ because these are equal
weight superpositions of the path marker states. Further, the
reduced state of the photon after the coupling has been
established is a mixture of the path states, so that any phase
relation between these states has been erased. Accordingly, the
detector outputs cannot detect any interference indicative of
coherence between the path input states, and the only information
left about the input is path information.

A third `marginal' input POVM is defined as follows:
\begin{eqnarray}\label{mz eqn:3.22}
H''_{1}=E''_{11}+E''_{22}&=&
\textstyle{\frac{1}{2}}\left(I-\sin\delta(\cos\gamma\,\sigma_x-\sin\gamma\,\sigma_y)\right)
\nonumber\\
H''_{2}=E''_{12}+E''_{21}&=&
\textstyle{\frac{1}{2}}\left(I+\sin\delta(\cos\gamma\sigma_x+\sin\gamma\,\sigma_y)\right).
\end{eqnarray}

This is a smeared interference observable, the unsharpness being
determined by the term $\sin\delta$ and the direction of the
associated Poincar\'e sphere vector being given by
$\pm(\cos\gamma,\sin\gamma,0)$. By varying $\gamma$ from 0 to
$2\pi$, all possible interference observables can be realized. The
erasure scheme presented here constitutes a joint unsharp
measurement of path and interference observables as represented by
the marginal POVMs $\{F''_1,F''_2\}$ and $\{H''_1,H''_2\}$.

We also see that for $\delta=-{\pi}/2$, all four effects
$E_{k\ell}$ are fractions of spectral projections of a sharp
interference observable; the marginal $\{H''_1,H''_2\}$ becomes a
sharp interference observable and the marginal $\{F''_1,F''_2\}$
becomes a trivial observable. Here we have recovered the
observation of SEW, that the detector statistics conditional on
the probe output readings display perfect interference with
perfect  visibility. In fact, we have found somewhat more:
independently of the photon input state, the conditional probabilities for
detections at $D_1$, $D_2$ given a probe recording of
$|q_1\rangle$ (say) are
\begin{eqnarray}\label{mz eqn:3.23}
prob(D_1|q_1)&=&\frac{\langle\psi_i|E''_{11}\psi_i\rangle}{\langle\psi_i|G''_1\psi_i\rangle}
=
\langle\psi_i|\textstyle{\frac 12}(I+\cos\gamma\,\sigma_x+\sin\gamma\,\sigma_y)\psi_i\rangle,\\
prob(D_2|q_1)&=&\frac{\langle\psi_i|E''_{21}\psi_i\rangle}{\langle\psi_i|G''_1\psi_i\rangle}
= \langle\psi_i|\textstyle{\frac
12}(I-\cos\gamma\,\sigma_x-\sin\gamma\,\sigma_y)\psi_i\rangle.
\end{eqnarray}
For $\gamma=0$ and the input state $\psi_i=\frac 1{\sqrt
2}(|1\rangle+|2\rangle)$, this gives $prob(D_1|q_1)=1$ and $prob(D_2|q_1)=0$.
This corresponds to the case of perfect interference fringes. Similarly,
for the detector probabilities  conditional on $|q_2\rangle$ and the above
input eigenstate of $\sigma_x$, we obtain probabilities $0$ and $1$ for D$_1$
and D$_2$, respectively, which are characteristic of interference antifringes.

This situation is a consequence of the fact that for the above
input and $\delta=- {\pi}/2$, $\gamma=0$, the state $\Psi_e$ and also the total output
state $\Psi_f^{-\pi/ 2}$ is an EPR state, analogous to the state described in the SEW
erasure setup of subsection \ref{sec:SEW}:
\begin{eqnarray}\label{mz eqn:3.24}
\Psi_f^{-{\pi}/2}&=&\textstyle{\frac{-(1-i )}{\sqrt
2}}\left[\textstyle{\frac{1}{\sqrt{2}}}(|1\rangle-|2\rangle)\otimes|p_1\rangle
+\textstyle{\frac{1}{\sqrt
2}}(|1\rangle+|2\rangle)\otimes|p_2\rangle\right]\nonumber
\\
&=&\textstyle{\frac{-(1-i )}{\sqrt 2}}\textstyle{\frac 1{\sqrt
2}}\left[|1\rangle\otimes|q_1\rangle+|2\rangle\otimes|q_2\rangle\right].
\end{eqnarray}

\subsection{Quantitative erasure}\label{sec:quant eras}

The two possible experimental options discussed in the preceding
subsections, of path marking and erasure are mutually exclusive in
that they require settings and operations that cannot be performed
at the same time: for path determination, the probe eigenstates
$|p_1\rangle,|p_2\rangle$ must be read out, while for the recovery
of interference it is necessary to record the detector outputs
conditional on the probe output states $|q_1\rangle,|q_2\rangle$.
Erasure was achieved by choosing $\delta=\frac{\pi}2$, which led
to the POVM $\{E''_{k\ell}\}$ being constituted of (fractions of)
spectral projections of an interference observable. Accordingly,
the only non-trivial marginal is the sharp interference observable
$\{H''_1,H''_2\}$.

If, however, the interferometric parameter $\delta$ is varied
between 0 and $\frac{\pi}2$, then the POVM $\{E''_{k\ell}\}$ is a
joint observable for an unsharp path and an unsharp interference
observable. In this case the experiment provides simultaneous
information about these noncommuting quantities. In the limit of
$\delta=0$, the interference marginal $\{H''_1,H''_2\}$ becomes
trivial and the path marginal $\{F''_1,F''_2\}$ becomes sharp.

The possibility of obtaining some joint information about both
observables, path and interference, can also be achieved by
modifying the path marking coupling in such a way that the
correlation between the paths and the probe indicator observable
is not perfect. This has been described as \emph{quantitative
erasure} (e.g., \cite{EnglertBergou2000}). We show here that
quantitative erasure is again an instance of a joint unsharp
measurement.

We take the path-marking interaction to be of the same form as
before, Eq.~(\ref{mz eqn:3.3}), but now we specify the marker
states $|p_1\rangle,|p_2\rangle$ to be nonorthogonal. Their
associated Poincar\'e sphere vectors will be chosen to be tilted
by an angle $\theta$ away from the $\pm z$ axis towards the
positive $x$ axis. As  pointer states we choose
$|q_1\rangle,|q_2\rangle$ equal to the up and down eigenstates of
$\sigma_z$. Thus we define
\begin{equation}\label{mz eqn:3.25}
|p_1\rangle=\textstyle{\cos\frac\theta 2 |q_1\rangle+\sin\frac\theta 2
|q_2\rangle},\ |p_2\rangle=\textstyle{\sin\frac\theta 2 |q_1
\rangle+\cos\frac\theta 2 |q_2\rangle}.
\end{equation}

The final state after the path-marking interaction is
\begin{eqnarray}\label{mz eqn:3.27}
\Psi_f^{\delta,\theta}&=&\left[\left(-\textstyle{\frac\alpha 2\cos\frac\theta
2}(1+e^{i\delta})+ i\textstyle{\frac\beta
2\sin\frac\theta 2}(1-e^{i\delta})\right)\,|1\rangle\right.\nonumber \\
 &\quad +&\left. \left(-i\textstyle{\frac\alpha 2\cos\frac\theta 2}(1-e^{i\delta})-
\textstyle{\frac\beta 2\sin\frac\theta 2}(1+e^{i\delta})\right)\,
|2\rangle\right]\otimes|q_1\rangle\nonumber\\
&\quad +&\left[\left(-\textstyle{\frac\alpha 2\sin\frac\theta
2}(1+e^{i\delta})+ i\textstyle{\frac\beta 2\cos\frac\theta
2}(1-e^{i\delta})\right)\,|1\rangle\right. \\
&\quad +&\left. \left(-i\textstyle{\frac\alpha 2\sin\frac\theta
2}(1-e^{i\delta})- \textstyle{\frac\beta 2\cos\frac\theta
2}(1+e^{i\delta})\right)\,|2\rangle\right]\otimes|q_2\rangle.\nonumber
\end{eqnarray}

Now we determine the input effects $E'''_{k\ell}$ associated with the
output projections $M'''_{k\ell}=|k\rangle\langle
k|\otimes|q_\ell\rangle\langle q_\ell |$ via
$\langle\Psi_f^{\delta,\theta}|M'''_{k\ell}|\Psi_f^{\delta,\theta}\rangle=
\langle\psi_i|E'''_{k\ell}|\psi_i\rangle$,
\begin{eqnarray}\label{mz eqn:3.28}
E'''_{11}&=&\textstyle{\frac{1}{4}}[I(1+\cos\theta\cos\delta)-\sin\delta\sin\theta\,\sigma_x
+(\cos\delta+\cos\theta)\,\sigma_z]\nonumber\\
&=&\textstyle{\frac{1}{4}}[I(1+\cos\theta\cos\delta)+\mathbf{m}'''\cdot\sigma]\nonumber\\
E'''_{21}&=&\textstyle{\frac{1}{4}}[I(1-\cos\theta\cos\delta)+\sin\delta\sin\theta\,\sigma_x
-(\cos\delta-\cos\theta)\,\sigma_z]\nonumber\\
&=&\textstyle{\frac{1}{4}}[I(1-\cos\theta\cos\delta)-\mathbf{n}'''\cdot\sigma]\\
E'''_{12}&=&\textstyle{\frac{1}{4}}[I(1-\cos\theta\cos\delta)-\sin\delta\sin\theta\,\sigma_x
+(\cos\delta-\cos\theta)\,\sigma_z]\nonumber\\
&=&\textstyle{\frac{1}{4}}[I(-\cos\theta\cos\delta)+\mathbf{n}'''\cdot\sigma]\nonumber\\
E'''_{22}&=&\textstyle{\frac{1}{4}}[I(1+\cos\theta\cos\delta)+\sin\delta\sin\theta\,\sigma_x
-(\cos\delta+\cos\theta)\,\sigma_z]\nonumber\\
&=&\textstyle{\frac{1}{4}}[I(1+\cos\theta\cos\delta)-\mathbf{m}'''\cdot\sigma],\nonumber
\end{eqnarray}
where
\begin{eqnarray}\label{mz eqn:3.29}
\mathbf{m}'''&=&(-\sin\delta\sin\theta,0,(\cos\delta+\cos\theta)),\nonumber\\
\mathbf{n}'''&=&(-\sin\delta\sin\theta,0,(\cos\delta-\cos\theta)).
\end{eqnarray}
The three marginal POVMs are determined as before:
\begin{eqnarray}\label{mz eqn:3.30}
F'''_{1}&=&E'''_{11}+E'''_{12}=
\textstyle{\frac{1}{2}}(I-\sin\delta\sin\theta\,\sigma_x+\cos\delta\,\sigma_z)\nonumber\\
F'''_{2}&=&E'''_{22}+E'''_{21}=
\textstyle{\frac{1}{2}}(I+\sin\delta\sin\theta\,\sigma_x-\cos\delta\,\sigma_z)\nonumber\\
G'''_{1}&=&E'''_{11}+E'''_{21}=
\textstyle{\frac{1}{2}}(I+\cos\theta\,\sigma_z)\nonumber\\
G'''_{2}&=&E'''_{22}+E'''_{12}=
\textstyle{\frac{1}{2}}(I-\cos\theta\,\sigma_z)\\
H'''_{1}&=&E'''_{11}+E'''_{22}=
\textstyle{\frac{1}{2}}I(1+\cos\theta\cos\delta)\nonumber\\
H'''_{2}&=&E'''_{12}+E'''_{21}=
\textstyle{\frac{1}{2}}I(1-\cos\theta\cos\delta).\nonumber
\end{eqnarray}
For $\delta=-{\pi}/2$, the first marginal POVM (corresponding
to the detector statistics) becomes an unsharp interference
observable, while the second marginal POVM (corresponding to the
probe output statistics) is an unsharp path observable. In both
cases  the unsharpness is determined by the parameter $\theta$.

We note for later reference that instead of the choice of pointer states $|q_1\rangle,|q_2\rangle$, one could have measured any pair of orthogonal probe states $|r_1\rangle,|r_2\rangle$. It is straightforward to show that the marker marginal is always an unsharp path observable.

\section{Complementarity and uncertainty in Mach-Zehnder interferometry}\label{sec:c-u}

\subsection{Manifestations of complementarity in Mach-Zehnder interferometry}\label{sec:c in mzi}

The sequence of experiments in section \ref{sec:MZ} is a demonstration of
complementarity in different guises. In the first two experiments path
detection (\ref{sec:none}) and interference detection (\ref{sec:phase}) are
mutually exclusive because this requires settings of the parameter $\delta$ which
cannot be realized in the same experiment, namely,  $\delta=0$ for path ($\sigma_z$)
measurement and $\delta=-{\pi}/2$ for interference ($\sigma_x$) measurement,
respectively. Here we have an instance of the complementarity of measurement setups or
measurement complementarity: these two noncommuting sharp observables do not
admit any joint measurement.

These experiments can also be used to confirm
preparation complementarity. We recall that sending a path eigenstate
$|1\rangle$ or $|2\rangle$ into the Mach-Zehnder interferometric setup to observe interference leads to
the probability
$\langle1|E_{1}^{-{\pi}/2}|1\rangle=\langle1|E_{2}^{-{\pi}/2}|1\rangle={1}/2$,
interference is completely uncertain. And, if we feed an interference
eigenstate, $|\pm x\rangle$, into the interferometer to measure path ($\delta=0$),
the path observable is uncertain $\langle\pm x|E_{1}^{0}|\pm x\rangle=\langle
\pm x|E_{2}^{0}|\pm x\rangle={1}/2$.

Value complementarity can indeed be used to explain the disappearance of
interference fringes resulting from path marking. Once perfect
correlation between the path states and the marker states is established in the
entangled state (\ref{mz eqn:3.3}), the reduced state of the photon is a
mixture of the path eigenstates $|1\rangle$ and $|2\rangle$. In each of these,
the path is definite, and therefore, in accordance with value complementarity,
the outcomes of a subsequent interference measurement are equally probable. No
interference fringes show up. Indeed this remains true for any mixture of path
eigenstates.

This account in terms of preparation complementarity views the path marking
interaction as part of a preparation process. An alternative explanation is
possible in terms of measurement complementarity as follows.

In the experiment of section \ref{sec:mark}, where sharp path marking is
followed by the interference setup with $\delta=-{\pi}/2$, it was found
that the path measurement interaction leads to a complete loss of interference
information detectable in D$_1$, D$_2$. All that the detectors can ``see'' is
path information, irrespective of the value of $\delta$ (Eq.~(\ref{mz
eqn:3.13})).

If the sharp path marking is relaxed into unsharp path marking, section
\ref{sec:quant eras}, setting the interferometer with $\delta=-{\pi}/2$
defines an unsharp  interference observable, which is jointly measured with the path
that can be recorded at the path marker.

It is found that the less accurate the path marking is set by making
$\cos\theta$ in $G'''_{1,2}=\frac 12(I\pm \cos\theta\,\sigma_z)$ smaller, the
sharper will be the interference measurement as $\sin\theta$ in
$F'''_{1,2}=\frac 12(I\mp\sin\theta\,\sigma_x)$ becomes larger.

We see here that measurement complementarity follows in the limits of making
the path marginal or the interference marginal perfectly sharp, rendering the
other trivial.

A similar analysis applies to the erasure setup (Sec.~\ref{sec:erasure}): if
$0<\delta<{\pi}/2$, this setup realizes a joint measurement of the POVMs
$\{F''_{1,},F''_2\}$ and $\{H''_{1,},H''_2\}$ which are unsharp path and
interference observables.

Consider the case of $\delta={\pi}/2$, where $F''_{1,2}=\frac{1}{2}I$ and
$H''_{1,2}=\frac{1}{2} (I\mp\cos\gamma\sigma_x\mp\sin\gamma\sigma_y)$. Here we
have a sharp interference measurement and no path measurement. With $\delta=0$,
$F''_{1,2}=\frac{1}{2}(I\pm\sigma_z)$ and $H''_{1}=\frac{1}{2}I$, so that we
have a sharp path measurement and no interference. These two limit cases of a
joint measurement scheme illustrate once more measurement complementarity.

\subsection{Value complementarity from preparation uncertainty relations}\label{sec:vc from pu}

We recall that a pair of value complementary observables $A,B$ is characterized
by the condition that in each of the eigenstates of $A$, all
eigenvalues of $B$ are equally likely to occur as outcomes of a
$B$ measurement; and vice versa.  We now
show that for qubit observables such as those occurring in the Mach-Zehnder
interferometry measurements discussed here, the value
complementarity property can always be obtained as a consequence
of some suitable uncertainty relation for the observables in
question.

In what follows we allow general states represented as density
operators $\rho=\frac 12(I+\mathbf{r}\cdot\sigma)$,
$|\mathbf{r}|\le 1$. For the observables represented by
$\sigma_x$, $\sigma_y$ and $\sigma_z$, we then have
\begin{equation}\label{expect-sigmas}
\langle\sigma_k\rangle=r_k^2,\quad k=1,2,3,
\end{equation}
and the variances, defined as
$\var(A)\equiv\var(A,\rho)=\langle A^2\rangle-\langle A\rangle^2$
(where $\langle A\rangle=\mathrm{tr}[A\rho]$, etc., $\mathrm{tr}[\cdot]$ 
denoting the trace operation), are
\begin{equation}\label{var-sigmas}
\var(\sigma_k)=1-\langle\sigma_k\rangle^2=1-r_k^2,\quad k=1,2,3.
\end{equation}
With these expressions it is straightforward to confirm the general
uncertainty relation for variances, with the commutator and covariance
terms contributing the the lower bound:
\begin{equation}\label{eqn:xz-ur}
\var(\sigma_x)\var(\sigma_z)\ \ge\ \textstyle{\frac
14}|\langle[\sigma_x,\sigma_z]\rangle|^2 + \textstyle{\frac
14}[\langle\sigma_x\sigma_z+\sigma_z\sigma_x\rangle-2\langle\sigma_x\rangle
\langle\sigma_z\rangle]^2.
\end{equation}

It is easy to verify that the pair $\sigma_x,\sigma_z$ is value complementary: in an eigenstate of
$\sigma_z$, we have $r_1^2=1$ and so $r_2=0$. So, probability 1 for a value of $\sigma_z$ goes
along with probability 1/2 for the values of $\sigma_x$; and vice versa.

However, when it comes to deciding whether the statement of value complementarity can be
inferred from the uncertainty principle, one should look at the uncertainty relation (\ref{eqn:xz-ur})
alone, rather than using the explicit values of its terms.
But using solely the above variance inequality one cannot recover value complementarity without
further information on the terms of the left hand side. Still it suffices to use the algebraic
and spectral properties of the Pauli operators (we imagine that we are given only this information
but not the explicit expression of the probabilities or expectations): then one finds the right hand side
of Eq.~(\ref{eqn:xz-ur}) to be equal to
$\langle\sigma_y\rangle^2+\langle\sigma_x\rangle^2\langle\sigma_z\rangle^2$,
and one can argue as follows: if the path is definite, that is, if
$\rho=|\psi\rangle\langle\psi|$ with $\psi$  an eigenstate of $\sigma_z$, the
left hand side of the uncertainty relation (\ref{eqn:xz-ur}) is zero, and
therefore the terms on the right hand side must vanish, too. Thus as
$\langle\sigma_z\rangle=1$, then
$\langle\sigma_x\rangle=\langle\sigma_y\rangle=0$. Since the eigenvalues of
these quantities are $\pm 1$, it follows that these observables are uniformly
distributed in $\psi$. By symmetry, $\sigma_z$ is uniformly distributed if
$\sigma_x$ has a definite value. (Note that we did not have to use the full
explicit expressions for the expectation values, which would of course make
this consideration entirely trivial.)

There are other forms of uncertainty relations which yield the statement of
value complementarity without recourse to specific properties of the Pauli
operators. Here we only mention the entropic uncertainty for two observables
$A,B$ with spectral representations $A=\sum_{i=1}^2a_iP_i$,
$B=\sum_{k=1}^2b_kQ_k$. The (Shannon) entropy of $A$ in a state $\psi$ is
defined as
\begin{equation}
H(A,\psi)=-\sum_{i=1}^{2}\langle\psi| P_i|\psi\rangle\log_2\langle\psi| P_i|\psi\rangle.
\end{equation}
This quantity is a measure of uncertainty concerning the value of $A$ as
encoded in the probability distribution of $A$ in the given state: note that
$0\le H(A,\psi)\le \log_2 2=1$, where the lower bound is assumed for any
eigenstate of $A$ and the upper bound arises for any state which assigns equal
probability $1/2$ to all eigenvalues. The following \emph{additive} trade-off
relation then holds \cite{Partha}:
\begin{equation}
H(A,\psi)+H(B,\psi)\ge -2\log_2\left(\max_{i,k}\frac{|\langle\psi|
P_iQ_k|\psi\rangle|}{\|P_i\psi\|\,\|Q_k\psi\|}\right).
\end{equation}
If $A,B$ are a pair of value complementary observables, so that any pair of
eigenstates $\psi_i$ of $A$ and $\phi_k$ of $B$ have overlap given by
$|\langle\psi_i|\phi_k\rangle|= 1/{\sqrt{2}}$, it follows that the lower bound
on the right hand side is $\log_22$. Thus, for $A=\sigma_z$, $B=\sigma_x$, one
obtains \cite{MaassenUffink1988}:
\begin{equation}\label{eqn:2-entro}
H(\sigma_z,\psi)+H(\sigma_x,\psi)\ge 1.
\end{equation}
The combined lack of information about $\sigma_z$ and
$\sigma_x$ is never less than one bit. Now, it is easily seen from
this inequality alone that if one observable has a definite value,
e.g., $H(\sigma_z,\psi)=1$ (which happens in the eigenstates of
$\sigma_z$), then the other observable is maximally uncertain, it
is uniformly distributed since $H(\sigma_x,\psi)=1$.

We note that the explicit expressions for the probabilities have gone into
the derivation of this entropic inequality. But all that is needed to recover the
property of value complementarity is contained in this inequality.

The inequality (\ref{eqn:2-entro}) is indeed easily verified by using the expressions
$\frac 12(1\pm r_z),$ $\frac 12(1\pm r_x)$ for the probabilities and applying
calculus to determine the minima. It is similarly straightforward to verify the
following \emph{triple} uncertainty relation:
\begin{equation}
H(\sigma_x,\psi)+H(\sigma_y,\psi)+H(\sigma_z,\psi)\ge 2.
\end{equation}

\subsection{Quantitative duality relations are uncertainty relations}

In the debates of the 1990s over complementarity in the context of
interferometry and which-path experiments, the meaning of the term
wave particle duality has gradually shifted away from a relation
of strict exclusion of path determination and interference
observation (in the same setup) to the broader idea of a
continuous trade-off between approximate path determination and
approximate interference determination. These discussions were eventually linked
with earlier theoretical and experimental work of the 1980s on simultaneous
but imperfect path determination and interference observation (e.g., \cite{WZ,MPS,Busch1987,GY}), 
as reviewed in \cite{DR2000}. The original intuitive ideas of the pioneers on an
approximate reconciliation of complementary operational options (cf.~ Sec.~\ref{sec:cu-rev}) 
have thus been turned into precise trade-off relations which are being tested experimentally.

Trade-off relations of the form
$P^2+V^2\le 1$ were derived as characterizations of the duality
between path predictability and interference visibility \cite{JHS,JS,JSV}. 
(In \cite{JSV}, a stronger result, $P^2+V^2=1$, was shown to hold for certain experimental situation.)
Soon afterwards, 
similar relations were formulated for quantum erasure (for a lucid
and comprehensive review, see \cite{EnglertBergou2000}). Vivid debates took
place over the question whether the associated quantities are connected with
uncertainties, and it has been shown that the respective trade-off relations are
related in various ways to some forms of uncertainty relations. In the context of the Mach-Zehnder
interferometer, we refer, in particular, to the work of Bj\"ork \emph{et al.}
\cite{Bjork1999}, D\"urr and Rempe \cite{DR2000}, and Luis \cite{Luis2001}.

Using familiar measures of uncertainty, we give a simple demonstration that in
the context of Mach-Zehnder interferometry experiments, quantitative duality relations are
indeed equivalent to  the uncertainty relation for an appropriate pair of associated
observables.

Any state $\rho$ can be represented by a matrix of the following
form in the basis of eigenvectors of $\sigma_z$:
\begin{equation}
\rho =\left(\begin{array}{cc}w_+&re^{-i\theta}\\
re^{i\theta}&w_-\end{array}\right),\quad\mathrm{where\ } \Bigg\{
\begin{array}{c}w_{\pm} \ge 0,\ w_++w_-=1,\\
 0 \le r\le \sqrt{w_+w_-},\ 0\le\theta< 2\pi.\end{array}
\end{equation}
We define the \emph{path contrast} of  $\rho$ as
\begin{equation}
C_P=C_P(\rho):=|prob(\sigma_z=+1,\rho)-prob(\sigma_z=-1,\rho)| =|w_+-w_-|.
\end{equation}
This is identical to the predictability $P$ entering the duality
relation $P^2+V^2\le 1$. Similarly we define the interference
contrast of $\rho$ as
\begin{equation}
C_I=C_I(\rho)=|prob(\sigma_x=+1,\rho)-prob(\sigma_x=-1,\rho)| =2r\cos\theta.
\end{equation}
With the specification $\theta=0$, or with an alternative choice of interference observable,
 this reduces to the visibility
$V:=(I_{max}-I_{min})/(I_{max}+I_{min})=2r$ (where $I_{max},I_{min}$ denote the maximal
and minimal intensities of the measured interference pattern, obtained by variation of
the interference observables). Using $r^2\le
w_+w_-$, the following duality relation is easily verified:
\begin{equation}
C_P^2+C_I^2=w_+^2+w_-^2-2w_+w_-+4r^2\cos^2\theta\le 1.
\end{equation}
Now we observe that
\begin{equation}
C_P^2=\langle\sigma_z\rangle^2=1-\var(\sigma_z),\quad
C_I^2=\langle\sigma_x\rangle^2=1-\var(\sigma_x).
\end{equation}
Therefore, the above duality inequality can be equivalently
expressed as
\begin{equation}
\var(\sigma_z)+\var(\sigma_x)=2-(C_P^2+C_I^2)\ge 1.
\end{equation}
Thus, our duality relation is equivalent to a form of uncertainty
trade-off relation. As before, value complementarity is again
entailed as a limit case.

We now show that this last inequality is actually a direct
consequence of the uncertainty relation (\ref{eqn:xz-ur}). Using Eq. (\ref{expect-sigmas}),
that relation is readily found to be equivalent to
\begin{equation}
\langle\sigma_x\rangle^2+\langle\sigma_y\rangle^2+\langle\sigma_z\rangle^2\le 1,
\end{equation}
which expresses the positivity of the state $\rho$.
Using Eq.~(\ref{var-sigmas}), we
thus see that the uncertainty relation (\ref{eqn:xz-ur}) is indeed
equivalent to the following inequality:
\begin{equation}\label{eqn:triple var}
\var(\sigma_x)+\var(\sigma_y)+\var(\sigma_z)\ge 2.
\end{equation}
We note that besides $\sigma_x$, the operator $\sigma_y$ also constitutes an
interference observable with respect to the path $\sigma_z$. Thus, substituting
$\var(\sigma_z)=1-C_P^2$, $\var(\sigma_x)=1-C_I^2\equiv1-C_{I,x}^2$, and a
similar term$\var(\sigma_y)=1-C_{I,y}^2$, we obtain a generalized and sharpened
duality relation which involves one path and two complementary interference
observables:
\begin{equation}\label{triple}
C_P^2+C_{I,x}^2+C_{I,y}^2\le 1.
\end{equation}
Thus the full uncertainty relation for $\sigma_x,\sigma_z$, including the
commutator and covariance terms, is equivalent to the additive triple trade-off
relation for the variances of $\sigma_x,\sigma_y,\sigma_z$ as well as this new
trade-off relation for three mutually complementary observables.

\subsection{Measurement complementarity from measurement inaccuracy relations}\label{mc from mir}

The measurement schemes of subsections \ref{sec:erasure} and  \ref{sec:quant eras} were found
to constitute joint measurements of unsharp path and interference
observables of the form $F=\{F_{1,2}=\frac 12(I\pm f\sigma_x\}$
and $G=\{G_{1,2}=\frac 12(I\pm g\sigma_z)$. For instance, in
equation (\ref{mz eqn:3.30}), setting $\delta=-\pi/ 2$, we
have $f=\sin\theta$ and $g=\cos\theta$, so that we have
$f^2+g^2=1$. This is an instance of  the criterion (\ref{joint-mt-condit})
which ensures the joint measurability of the POVMs $F,G$.

For a general state $\rho=\frac 12(I+\mathbf{r}\cdot\sigma)$,
$|\mathbf{r}|\le 1$, we define the contrasts of the distributions
of $F$ and $G$,
\begin{eqnarray}
C_F(\rho)&=&|\tr[\rho F_1]-\tr[\rho F_2])|=|fr_1|,\nonumber\\
C_G(\rho)&=&|\tr[\rho G_1]-\tr[\rho G_2]|=|gr_3|.
\end{eqnarray}
The contrasts of the POVMs $F,G$ are the respective maximal
contrasts over all states $\rho$:
\begin{equation}
C_F=|f|,\quad C_G=|g|.
\end{equation}
These quantities measure the degrees of unsharpness,
\begin{equation}
U_F:=1-C_F^2=1-f^2,\quad U_G:=1-C_G^2=1-g^2,
\end{equation}
in the POVMs $F,G$. The unsharpness of $F$ can also be defined as
the minimum variance of the distribution of $F$ for all states
$\rho$. Indeed, it is not hard to verify that
\begin{eqnarray}\label{eqn:var povm}
\var_\rho(F)=1-f^2r_1^2&=&1-f^2+f^2(1-r_1^2)\nonumber\\
&=& U_F+(1-U_F)\var_\rho(\sigma_x)\ge U_F.
\end{eqnarray}
Taking the minimum over all $\rho$ gives
\begin{equation}
\var_{min}(F)=U_F.
\end{equation}
The above joint measurability criterion can be written in terms of
the degrees of unsharpness:
\begin{equation}
U_F+U_G\ge 1.
\end{equation}
This inequality is an uncertainty trade-off relation which must be
satisfied if the two noncommuting unsharp path and interference
observables $F$ and $G$ are to be jointly measurable. We have here
an instance of Heisenberg's uncertainty principle for the
inaccuracies which are necessarily present in joint measurements.
As far as we are aware, this is one \cite{BuShill2003} of two cases in which an
inaccuracy relation has been proven as a necessary condition for
joint measurability. The other example is the case of position and
momentum \cite{Carmeli,Wer}, and the corresponding uncertainty relation for joint
measurements is reviewed in \cite{BuHeiLah2006}.

Finally we note that the variances of the marginals $F,G$ in a
joint measurement satisfy the uncertainty relation
\begin{equation}
\var_\rho(F)+\var_\rho(G)\ge U_F+U_G\ge 1.
\end{equation}

Measurement complementarity is obtained as a limiting case for a
pair $F,G$ which are jointly measurable: if it is stipulated that
one marginal, say $F$, becomes sharp,  $U_F=0$, or $|f|=1$, then
the other marginal, $G$, becomes a trivial POVM, $g=0$,
$G_{1,2}=\frac 12 I$. Thus, if the path $F$ is measured sharply,
any attempt at obtaining information on interference will fail as
the only unsharp interference observable $G$ that can be measured
jointly with $F$ is trivial.

\subsection{Disturbance versus accuracy}

The setup discussed in Subsection \ref{sec:quant eras} corresponds to a
sequence of measurements where first path marking and registration can be
achieved and then an interferometric measurement is carried out. As was
observed in Subsection \ref{sec:c in mzi}, if the path marking correlation is
perfect, all the interferometric detectors can ``see" is unsharp path
information. If the setting is $\delta=\pi/ 2$, then the input observable
indicated by the detectors is trivial: no interference information whatsoever
about the input state is observed. If the path marking is unsharp, due to
imperfect correlations or nonorthogonal path marker states
$|p_1\rangle,|p_2\rangle$, then \emph{some} interference information about the
input state can pass through the path marking interaction. Here we will give a
quantitative expression of this trade-off between the ``disturbance" of the
interference information through (imperfect) path marking and the accuracy of
this path-marking process.

The path-marking interaction transforms the initial state
$(\alpha|1\rangle+\beta|2\rangle)|p_0\rangle$ into
\begin{equation}
\Psi_e:=\alpha|1\rangle|p_1\rangle+\beta|2\rangle|p_2\rangle.
\end{equation}
This state has reduced photon-state operator
\begin{equation}
\rho_e 
=|\alpha|^2|1\rangle\langle 1|+|\beta|^2|2\rangle\langle 2| \
+\alpha\beta^*\langle p_2|p_1\rangle|1\rangle\langle 2|+ \alpha^*\beta\langle
p_1|p_2\rangle|2\rangle\langle 1|.
\end{equation}
In what follows we want to express the necessary ``disturbance" of the
interference detection due to the path-marking entanglement. We start with a
situation where $|p_1\rangle,|p_2\rangle$ are not necessarily orthogonal, but
arranged as described in subsection \ref{sec:quant eras}. We also allow a
general set of orthogonal pointer states $|r_1\rangle,|r_2\rangle$.

The quality of the path marking can be determined by following the entangling
interaction with a path detection at D$_1$, D$_2$, where $\delta=0$. If a
reading $r_1$ ($r_2$) is taken to infer the path to be path 1 (path 2), then
the subsequent path detection in the interferometer can be used to verify this
inference. Thus, a joint measurement is made of the PVM with projections
$M_{k\ell}=|k\rangle\langle k|\otimes |\ell\rangle\langle\ell|$, with
probabilities
\begin{equation}
p_{k\ell}=\langle\Psi_f^0|M_{k\ell}|\Psi_f^0\rangle=|\langle k|\langle\ell|\Psi_e\rangle|^2.
\end{equation}
The probability $prob(corr)$ of a correct inference is given as the sum of the
probabilities of the corresponding coincident outputs:
\begin{equation}
prob(corr)=p_{11}+p_{22}=\langle\psi_i|H_1^0|\psi_i\rangle\ =|\alpha|^2|\langle
r_1|p_1\rangle|^2+|\beta|^2|\langle r_2|p_2\rangle|^2.
\end{equation}
Similarly, the probability $prob(err)$ of error is given by the non-coincident combinations:
\begin{equation}
prob(err)=p_{12}+p_{21}=\langle\psi_i|H_{2}^0|\psi_i\rangle\
=|\alpha|^2|\langle r_2|p_1\rangle|^2+|\beta|^2|\langle r_1|p_2\rangle|^2.
\end{equation}
Here we have introduced the input marginal $H^0=\{H_1^0,H_2^0\}$ which
represents the coincidence events in this joint measurement:
\begin{eqnarray}
H_1^0&=&|1\rangle\langle 1| |\langle r_1|p_1\rangle|^2+|2\rangle\langle 2|
|\langle r_2|p_2\rangle|^2,\nonumber\\
H_2^0&=&|1\rangle\langle 1| |\langle r_2|p_1\rangle|^2+|2\rangle\langle 2|
|\langle r_1|p_2\rangle|^2.
\end{eqnarray}
On writing $|r_m\rangle\langle r_m|=\frac 12(I+\mathbf{r}_m\cdot\sigma)$, $\mathbf{r}_1=\mathbf{r}$,
$\mathbf{r}_2=-\mathbf{r}$, and $|p_\ell\rangle\langle p_\ell|=\frac 12(I+\mathbf{p}_\ell\cdot\sigma)$,
we obtain the following forms for $H_1^0,H_2^0$:
\begin{eqnarray}
H_1^0&=\textstyle{\frac 12}\left(I(1+\textstyle{\frac 12}\mathbf{r}\cdot(\mathbf{p}_1-\mathbf{p}_2))
+\textstyle{\frac 12}\mathbf{r}\cdot(\mathbf{p}_1+\mathbf{p}_2)\sigma_z\right),\\
H_2^0&=\textstyle{\frac 12}\left(I(1-\textstyle{\frac
12}\mathbf{r}\cdot(\mathbf{p}_1-\mathbf{p}_2)) -\textstyle{\frac
12}\mathbf{r}\cdot(\mathbf{p}_1+\mathbf{p}_2)\sigma_z\right).
\end{eqnarray}
in this notation the probability $prob(corr)$ becomes
\begin{equation}
prob(corr)=\textstyle{\frac 12}\left(1+\mathbf{r}\cdot(|\alpha|^2\mathbf{p}_1-|\beta|^2\mathbf{p}_2)\right).
\end{equation}

In order to determine the maximum quality path determination available by a
suitable choice of the path marker output observable (with eigenstates
$|r_1\rangle,|r_2\rangle$), we maximize the probability $prob(corr)$ of correct
inferences on the path from the registrations of $r_1,r_2$. The maximum
$prob_{max}(corr)$ is obained at
\begin{equation}
\mathbf{r}=\mathbf{r}^0=\frac{|\alpha|^2\mathbf{p}_1-|\beta|^2\mathbf{p}_2}
{\left| |\alpha|^2\mathbf{p}_1-|\beta|^2\mathbf{p}_2\right|},
\end{equation}
and its value is
\begin{equation}\label{eqn:max corr}
prob_{max}(corr)=\textstyle{\frac 12}(1+\left||\alpha|^2\mathbf{p}_1-|\beta|^2\mathbf{p}_2\right|)=:L.
\end{equation}
The (path) distinguishability is defined as $D=2L-1$ \cite{JS}. We obtain:
\begin{equation}\label{disting}
D=\left||\alpha|^2\mathbf{p}_1-|\beta|^2\mathbf{p}_2\right|
=\sqrt{1-4|\alpha|^2|\beta|^2|\langle p_1|p_2\rangle|^2}.
\end{equation}
We note that the probabilities for correct and wrong inferences can be
expressed in terms of the coincidence POVM $H^0=\{H_1^0,H_2^0\}$: noting that
$Prob_{min}(err)=1-prob_{max}(corr)=1-L$, we have
\begin{equation}
D=\left[\langle\psi_i|H_1^0|\psi_i\rangle-
\langle\psi_i|H_2^0|\psi_i\rangle\right]_{\mathbf{r}=\mathbf{r}^0}.
\end{equation}
We calculate the variance of the distribution of $H^0$ in the state $\psi_i$,  using
\begin{equation}
\bar{t}=\int t\,d\langle\psi_i|H^0_t|\psi_i\rangle=
\langle\psi_i|H_1^0|\psi_i\rangle-\langle\psi_i|H_2^0|\psi_i\rangle,
\end{equation}
we obtain
\begin{equation}
\var(H,\psi_i)=\int (t-\bar{t})^2\,d\langle\psi_i|H^0_t|\psi_i\rangle\
=1-\left[\langle\psi_i|H_1^0|\psi_i\rangle-\langle\psi_i|H_2^0|\psi_i\rangle\right]^2.
\end{equation}
Thus we obtain that
\begin{equation}
D^2=1-\var(H^0,\psi_i)|_{\mathbf{r}=\mathbf{r}_0}.
\end{equation}
The distinguishability is thus found to be directly related to the uncertainty
of the coincidence observable. It gives a measure of the accuracy of the path
determination, and it is dependent on the degree of entanglement between the
photon and the marker system, which may be quantified by the overlap quantity $|\langle
p_1|p_2\rangle|$.

We now determine the visibility $V_e$ of an interference observation available
in the reduced photon state $\rho_e$ after the entangling interaction. Thus we
consider a joint measurement of the kind studied in subsection \ref{sec:quant
eras}. The usual definition of visibility in terms of the difference between
maximal and minimal probability of an outcome at D$_1$ (or D$_2$) reduces to
\begin{equation}
V_e=V(\rho_e)=\left|\tr(\rho_e|+,\mathbf{n}\rangle\langle+,\mathbf{n}|)-
\tr(\rho_e|-,\mathbf{n}\rangle\langle-,\mathbf{n}|)\right|,
\end{equation}
where the interference observable
\begin{equation}
\sigma_{\mathbf{n}}=|+,\mathbf{n}\rangle\langle+,\mathbf{n}|-
|-,\mathbf{n}\rangle\langle-,\mathbf{n}|=\mathbf{n}\cdot\sigma
=\left(\begin{array}{cc} 0&e^{-i \delta}\\ e^{i \delta}&0 \end{array}\right),
\end{equation}
(with $\mathbf{n}=(\cos\delta,\sin\delta,0)$) is chosen such that the above
difference becomes maximal:
\begin{equation}
V_e=2|\mathrm{Re}(\alpha\beta^*\langle p_2|p_1\rangle\,e^{i \delta})|_{max}=
2|\alpha||\beta||\langle p_2|p_1\rangle|.
\end{equation}
Since $V_e=|\langle\sigma_{\mathbf{n}}\rangle|$, we have
\begin{equation}
V_e^2=V(\rho_e)=1-\var(\sigma_{\mathbf{n}},\rho_e),
\end{equation}
and finally \cite{JSV}
\begin{equation}
V_e^2+D^2=1.
\end{equation}
This is a limiting case of a general inequality $V_e^2+D^2\le 1$, reviewed 
in \cite{EnglertBergou2000}, where it is shown that equality arises whenever the total system
of photon plus marker is in a pure state as is the case in the present context.

The above quantitative erasure duality relation can be written in terms of the
associated uncertainties:
\begin{equation}\label{eqn:disturb-ur}
\var(H^0,\psi_i)+\var(\sigma_{\mathbf{n}},\rho_e)=1.
\end{equation}
This relation shows the trade-off that happens as a result of the path-marking
interaction: the better the path marking is set, the more the interference term
becomes attenuated. This is to be understood as a relation between the accuracy
of path determination and the ``disturbance" of the interference capability of
the quantum state that passes through the path marking interaction. This can be
seen particularly clearly if the input state $\psi_i$ is chosen to be
an interference eigenstate, $|\alpha|=|\beta|=1/\sqrt{2}$. In this case the
distinguishability becomes minimal among all input states,  and the visibility
becomes maximal, and both quantities depend solely on the overlap between the
marker states and hence the degree of entanglement between photon and marker:
\begin{equation}\label{eqn:disturb}
D=|\mathbf{p}_1-\mathbf{p}_2|=\sqrt{1-|\langle p_1|p_2\rangle|^2},\
V_e=|\langle p_1|p_2\rangle|.
\end{equation}
The deviation of $V_e$ from 1, its original value for the input interference
eigenstate, represents the minimal degradation of the interference capability
required by the gain in path distinguishability. If the path marking is made
perfect, by requiring $\langle p_1|p_2\rangle=0$, then $D=1$ and $V_e=0$, that
is, the disturbance of the interference observation becomes maximal, no
interference can be detected.

\subsection{Uncertainty, disturbance, or entanglement?}

We finally turn to the question of whether entanglement provides a more
fundamental or more general explanation of the loss of interference in 
path-marking experiments than uncertainty. We think that this question arose from
the erroneous conflation of the uncertainty principle with the idea of
classical random disturbance. The discussion of that conflation has led to some
very interesting clarifications and distinctions between classical, random
momentum transfers (or phase kicks) and quantum-mechanical momentum transfers
(or phase shifts) \cite{Wiseman-etal,ESW2000}, with the conclusion that often,
if not always, one and the same experiment may admit explanations of the loss
of interference both in terms of classical random disturbances and in terms of
quantum disturbances.

It seems to be a desire for causal explanation that induced the search for
mechanical causes (classical or quantum) enforcing the uncertainty principle.
In the preceding subsection we have seen that the reverse perspective leads to
a satisfactory account of the disturbance of interference through path marking:
this disturbance is expressed by means of an uncertainty relation.

We have used the term ``disturbance" in the operational sense of a change in
the distribution of the values of the interference observable. There was no
question of a (random) mechanical kick. Quite  the contrary, the coupling for
the path marking process was arranged so as to constitute a
\emph{non-demolition} measurement: if the input is a path eigenstate, say
$|1\rangle$, the total state after path marking is $|1\rangle|p_1\rangle$; that
is, the system has remained undisturbed, it is still in the original path
eigenstate.

The magnitude of the loss of interference capability or coherence that arises
in the transition from an interference eigenstate as input to the reduced
photon state $\rho_e$ after the path marking interaction is determined solely
by the overlap $c:=|\langle p_1|p_2\rangle|$, Eq.~(\ref{eqn:disturb}). The
quantity $c$ describes the degree of correlation between the path eigenstates
and the marker states, as well as the degree of entanglement in the total state
$\Psi_e$.

It is evident that in the present quantitative erasure experiment, the
explanations of the degradation of coherence either in terms of uncertainties
for the photon system alone (Eq.~(\ref{eqn:disturb-ur})) or in terms of
entanglement draw on the same crucial entity --- the quantity $c$. The deeper
reason for this may be seen in the fact that entanglement can itself be
explained as an instance of uncertainty. This was made manifest in  the papers
of Kim and Mahler \cite{KM2000} and Bj\"ork \emph{et al.} \cite{Bjork1999}, who
used uncertainty relations for observables of the entangled object and probe
system to explain the loss of interference.

A pure state of a compound system is \emph{entangled} if it cannot be
represented as a product state. Product states are also called
\emph{separable}. The (normalized) state
$\Psi_e=\alpha|1\rangle|p_1\rangle+\beta|2\rangle|p_2\rangle$ is entangled
exactly when $c=|\langle p_1|p_2\rangle|<1$ and $0<|\alpha|,|\beta|<1$. As an
entangled state, $\Psi_e$ possesses a biorthogonal decomposition
$\Psi_e=\sqrt{w}\psi_1\otimes \phi_1+ \sqrt{1-w}\psi_2\otimes\phi_2$, where
$\langle\psi_1|\psi_2\rangle=0=\langle\phi_1|\phi_2\rangle$. The only values of
the (non-negative) parameter $w$ for which $\Psi_e$ is separable are $w=1$ or
$w=0$. Then the state $\Psi_e$ is separable exactly when the adapted observable
$S=|\psi_1\rangle\langle\psi_1|\otimes|\phi_1\rangle\langle\phi_1|-
|\psi_2\rangle\langle\psi_2|\otimes|\phi_2\rangle\langle\phi_2|$
has a definite value. Conversely, the state $\Psi_e$ is entangled if and only
if the outcomes of an $S$ measurement are uncertain. Seen in this way,
entanglement is quite generally an instance of uncertainty, and like
uncertainty, it is rooted in the existence of states which are superpositions
of orthogonal families of states. It is therefore to be expected that whenever
there is an explanation for a loss of interference due to entanglement, there
is also an associated explanation in terms of an uncertainty relation. Our
consideration of the previous subsection shows that such uncertainty relations
can even be formulated without recourse to the probe system.

It has been argued that in the case of the experiments of SEW and D\"urr
\emph{et al.}, the position-momentum uncertainty relation does not provide an
explanation of the loss of interference, and that no explanation based on
another uncertainty relation has been found \cite{DR2000}. However, the loss of
interference capability in the SEW double-slit experiment, discussed in Section
\ref{sec:SEW} above, is due to the loss of coherence in the transition  from
$\psi_0= \frac 1{\sqrt{2}}[\psi_1+\psi_2]$ to the associated mixed state,
\begin{equation}
|\psi_0\rangle\langle\psi_0|\longrightarrow \textstyle{\frac 12} |\psi_1\rangle\langle\psi_1|+
\textstyle{\frac 12} |\psi_2\rangle\langle\psi_2|,
\end{equation}
which is completely analogous to the transition
$|\psi_i\rangle\langle\psi_i|\to \rho_e$ of the photon state in the
Mach-Zehnder context. All that matters for the explanation of the loss of
interference is the presence or absence of a definite phase relation between
the path states $\psi_1$ and $\psi_2$ (or $| 1\rangle,|2\rangle$). This can be
fully described by analogs of Mach-Zehnder interferometric observables in the
two dimensional subspace spanned by the path states $\psi_1,\psi_2$. Thus if
the path observable is represented as
\begin{equation}
\Sigma_z:=|\psi_1\rangle\langle\psi_1|-|\psi_2\rangle\langle\psi_2|,
\end{equation}
then the phase relation between the path states can be tested by a phase
sensitive observable such as
\begin{equation}
\Sigma_x:=\textstyle{\frac 12}|\psi_1+\psi_2\rangle\langle\psi_1+\psi_2|-
\textstyle{\frac 12}|\psi_1-\psi_2\rangle\langle\psi_1-\psi_2|.
\end{equation}
The explanations and formulations of loss of interference in terms of
uncertainty relations given in the context of Mach-Zehnder interferometry in
the present section can then be literally transferred to the case of double-slit
interference and path-marking experiments.

The same conclusion has been drawn by Luis \cite{Luis2001}  who based his
argument on a continuous phase POVM conjugate to the path observable ($\sigma_z$),
instead of the phase-sensitive interference observables used in the present paper.
Similarly, Bj\"ork \emph{et al.} \cite{Bjork1999} and de Muynck and
Hendrikx \cite{deMHen2001} have demonstrated that analogous
atom-interferometric experiments can be interpreted in terms of joint unsharp
measurements of path and interference observables, and that complementarity
emerges as a limiting case of an entropic uncertainty relation for the
accuracies of these joint measurements.

\section{Conclusion}\label{sec:conc}

We have reviewed the evolution of the understanding and current formalizations
of the concepts of complementarity and uncertainty. In particular, we exhibited
three distinct versions of complementarity and uncertainty relations,
respectively, referring to the limitations and possibilities of preparing and
measuring simultaneously sharp or unsharp values, and to a necessary trade-off
between accuracy and state disturbance.

It was shown that  that these formulations can be usefully applied in the
context of Mach-Zehnder interferometry, to explain the loss or degradation of
interference due to path marking. In particular, it was found that value
complementarity and measurement complementarity can be recovered from
appropriate uncertainty relations for preparations, joint, and sequential
measurements, respectively. Furthermore, duality relations for the trade-off
between partial path determinations and reduced-visibility interference
observations were shown to be expressible as uncertainty relations.

Next, we have noted that entanglement is properly understood as an
instance of uncertainty in the context of the description of
compound systems, rather than a separate feature. It is therefore
to be expected that whenever an explanation of the loss of
interference can be given in terms of entanglement, this can be
accompanied with an explanation in terms of a form of uncertainty
relation.

We have pointed out  that ``disturbance" in quantum
measurements cannot be reduced to the naive notion of classical
(random) kicks; the generic concept of disturbance concerns the
necessary state change induced by measurements, which inevitably goes
along with the build-up of entanglement between object system and probe.
This applies, in particular, to the non-demolition measurement couplings
employed here for path marking, which are exactly of the same kind as
the coupling used in the SEW and D\"urr \emph{et al.} experiments.
Such (approximately) repeatable measurements were also at the heart of
Heisenberg's thought experiments illustrating the various types of uncertainty
relations. The necessary state change required for the extraction of path information
leads to a disturbance of the distribution of an associated interference observable.

All these notions --- uncertainty, complementarity, entanglement, and
disturbance --- are ultimately rooted in the linear structure of quantum
mechanics, with the concomitant noncommutativity of observables.

Taken all this together, it seems indeed moot to try and establish a hierarchy
of principles of uncertainty, complementarity, or entanglement \emph{within}
quantum mechanics. As seen from within this theory, these features are linked
with each other but cannot be claimed to be reducible to one another. They are
not logically independent, nor simply consequences of each other. Such logical
relations can only be analyzed within a framework more general than quantum
mechanics, in which each principle can be introduced as a separate postulate.
In fact, it has been shown that there are theories in which there is an
uncertainty principle without a complementarity principle (suitably specified),
and theories with complementarity but without a Heisenberg uncertainty relation
\cite{Lahti}. In quantum mechanics, uncertainty and complementarity are
consequences of the formalism which are neither reducible to each other nor
entirely unrelated.

There is thus no need  in quantum mechanics to speak of a complementarity or
uncertainty ``principle", unless one uses this term informally as a way of
highlighting these ``principal" implications of the theory, fundamental for its
intuitive understanding. If, however, one sets out to use complementarity or
uncertainty as \emph{postulates}, from which to deduce quantum mechanics in
Hilbert space starting from a more general framework, then it is indeed
appropriate to refer to them as principles. In this spirit, Bohr
\cite{Bohr1928} and Pauli \cite{Pauli1933} have referred to quantum mechanics
as ``the theory of complementarity", and the programme implied by this slogan
has been carried out in quite different ways by P.~Lahti together with the late
S.~Bugajski \cite{LahtiBug1985} (who use the convexity, or operational
approach) and by J.~Schwinger \cite{Schwinger} (based on his idea of
measurement algebra). Similarly compelling derivations of quantum mechanics
from the uncertainty principle, as envisaged by Heisenberg \cite{Heis1927}, are
still outstanding. This task seems impossible as long as the uncertainty
principle is expressed only in terms of an inequality; however, if the
uncertainty principle is reinterpreted in the stronger formal sense of the
canonical commutation relations, which represent the characteristic shift
covariance properties of localization observables, then a derivation of quantum
mechanics from such an enhanced postulate becomes possible starting, for
example, within a quantum logical framework \cite{Gudder1973}.

\section*{Acknowledgement} We are indebted to James Brooke (Saskatoon), Joy
Christian (Oxford and PI), Gregg Jaeger (Boston), Pekka Lahti (Turku), Abner Shimony (Boston) and Stefan Weigert (York) for
valuable comments and suggestions on various draft versions of this work. Part of
this work was carried out during P.~Busch's stay at Perimeter Institute.
C.~Shilladay gratefully acknowledges support and hospitality extended to him
during his short visit at PI.


\begin{thebibliography}{99}

\bibitem{SEW1991}  M.O.~Scully, B.-G.~Englert, H.~Walther, \emph{Quantum Optical
Tests of Complementarity}, Nature \textbf{351}, 111-116, 1991.

\bibitem{STCW1994}  P.~Storey, S.~Tan, M.~Collett, D.~Walls, \emph{Letter: Path
detection and the uncertainty principle} Nature, \textbf{367}, 626, 1994.

\bibitem{SEW1995} B-G.~Englert, M.~O.~Scully, H.~Walther, \emph{Letter: Complementarity
and uncertainty}, Nature \textbf{375}, 367, 1995.

\bibitem{DNR1998}  S.~D\"{u}rr, T.~Nonn, G.~Rempe, \emph{Origin of quantum-mechanical
complementarity probed by a `which-way' experiment in an atom interferometer},
Nature \textbf{395}, 33-37, 1998; \emph{Fringe Visibility and Which-Way Information
in an Atom Interferometer}, Physical Review Letters \textbf{81}, 5706-5709, 1998.

\bibitem{Buc1999}  M.~Buchanan, \emph{An end to uncertainty}, New Scientist, 6 March 1999,
 25-28.

\bibitem{Wiseman-etal} H.M.~Wiseman, F.E.~Harrrison, M.J.~Collett, S.M.~Tan, D.F.~Walls,
R.B.~Killip, \emph{Nonlocal momentum transfer in welcher Weg measurements},
Physical Review A \textbf{56}, 55-74, 1997.

\bibitem{LuisS-S} A.~Luis, L.L.~S\'anchez-Soto, \emph{Complementarity Enforced
by Random Classical Phase Kicks}, Physical Review Letters \textbf{81}, 4031-4035, 1998.

\bibitem{ESW2000} B.-G.~Englert, M.O.~Scully, H.~Walther, \emph{On mechanisms that
enforce complementarity}, Journal of Modern Optics \textbf{47}, 2213-2220, 2000.

\bibitem{Bjork1999}  G.~Bj\"{o}rk, J.~S\"{o}derholm, A.~Trifonov, T.~Tsegaye,
\emph{Complementarity and uncertainty relations},
Physical Review A \textbf{60}, 1874-1882, 1999.

\bibitem{KM2000}  I.~Kim, G.~Mahler, \emph{Uncertainty rescued: Bohr's
complementarity for a composite system},
Physics Letters A \textbf{269}, 287-292, 2000.

\bibitem{DR2000}  S.~D\"{u}rr, G.~Rempe, \emph{Can wave-particle duality be
based on the uncertainty relation?}  American Journal of Physics  \textbf{68}(11), 1021-1024, 2000.

\bibitem{Luis2001} A.~Luis, \emph{Complementarity and certainty relations for
two-dimensional systems}, Physical Review A \textbf{64}, 012103, 1-6, 2001.


\bibitem{Brandt1999} H.~E.~Brandt, \emph{Positive operator valued measures in quantum
information processing}, American Journal of Physics \textbf{67}(5) 434-439,
1999.

\bibitem{Peres} A.~Peres, \emph{Quantum Theory: Concepts and Methods}, Kluwer, Dordrecht, 1993.

\bibitem{Bohr1928}  N.~Bohr, \emph{The quantum postulate and the recent
development of atomic theory}, Nature \textbf{121}, 580-590
(1928).

\bibitem{Heis1927}  W.~Heisenberg, \emph{\"Uber den anschaulichen Inhalt der
quantentheoretischen Kinematik und Dynamik}, Zeitschrift f\"ur
Physik \textbf{43}, (1927), 172-198. English translation in
J.A.~Wheeler and W.Z.~Zurek, \emph{Quantum Theory and
Measurement}, Princeton University Press, 1983.

\bibitem{Jammer} M.~Jammer, \emph{The Philosophy of Quantum Mechanics},
John Wiley \&\ Sons, New York, 1974.

\bibitem{Heis1977} W.~Heisenberg, \emph{Remarks on the Origin of the Relations of Uncertainty},
in \emph{The Uncertainty Principle and Foundations of Quantum Mechanics --
A Fifty YearsÕ Survey}, eds. W.C.~Price, F.R.S., S.S.~Chissick. J.~Wiley \& Sons, London, 1977, pp.~3-6.

\bibitem{Pauli1933} W.~Pauli, \emph{General Principles of Quantum Mechanics},
Springer-Verlag, Berlin, 1980. (German original in Handbuch der Physik, 1933.)

\bibitem{Schwinger} J.~Schwinger, \emph{Quantum Kinematics and Dynamics},
W.A.~Benjamin, New York, 1970. See also: \emph{Quantum Mechanics: Symbolism of
Atomic Measurements}, Springer, Berlin, 2001.

\bibitem{Lahti1983} P.J.~Lahti, \emph{Hilbertian quantum theory as the theory of
complementarity}, International Journal of Theoretical Physics \textbf{22},
911-929, 1983.

\bibitem{Howard} D.~Howard, Who invented the Copenhagen Interpretation?
Philosophy of Science \textbf{71}, 669-682, 2004.

\bibitem{EPR1935} A.~Einstein, B.~Podolsky, N.~Rosen,
\emph{Can quantum-mechanical description of physical reality be considered complete?},
Physical Review \textbf{47}, 777-780, 1935.

\bibitem{Bohr1935} N.~Bohr, \emph{Can quantum-mechanical description of physical
reality be considered complete?}, Physical Review \textbf{48}, 696-702, 1935.

\bibitem{Heis1930} W.~Heisenberg, \emph{The Physical Principles of the Quantum Theory},
University of Chicago Press, 1930.

\bibitem{Cam2006} K.~Camilleri, \emph{Heisenberg and the wave-particle duality}, Studies
in History and Philosophy of Modern Physics \textbf{37}, 298-315, 2006.

\bibitem{BGL1995} P.~Busch, M.~Grabowski and P.~Lahti, \emph{Operational Quantum
Physics},  Springer-Verlag, Berlin, 1995, 2nd corrected printing 1997.

\bibitem{BuschLahti1995} P.~Busch, P.J.~Lahti, \emph{The Complementarity of Quantum Observables:
Theory and Experiments}, Rivista del Nuovo Cimento \textbf{18}, 1-27,1995.

\bibitem{Kraus1987} K.~Kraus, \emph{Complementary observables and uncertainty relations},
Physical Review D \textbf{35}, 3070-3075, 1987.

\bibitem{BuLahPel} P.~Busch, P.J.~Lahti, J.P.~Pellonp\"a\"a, K.~Ylinen,
\emph{Are number and phase complementary observables?},
Journal of Physics A \textbf{34}, 5923-5935, 2001.


\bibitem{vN1932} J.~von Neumann, \emph{Mathematische Grundlagen der
Quantenmechanik}, Springer-Verlag, Berlin, 1932. Reprinted 1980. English
translation: \emph{Mathematical Foundations of Quantum Mechanics}, Princeton
University Press, 1955.

\bibitem{Ozawa} M.~Ozawa, \emph{Universally valid reformulation of the Heisenberg uncertainty
principle on noise and disturbance}, Physical Review A \textbf{67}, 042105, 1-6, 2003.

\bibitem{Wer} R.F.~Werner, \emph{The Uncertainty Relation for Joint Measurement
of Position and Momentum}, Quantum Information and Computation \textbf{4}, 546-562, 2004.

\bibitem{BuHeiLah2006} P.~Busch, T.~Heinonen, P.J.~Lahti,, \emph{Heisenberg's
Uncertainty Principle: Three Faces, Two R\^oles}, forthcoming, 2006.

\bibitem{LahtiMac} P.J.~Lahti, M.J.~Maczynski, \emph{Heisenberg inequality and the complex 
field in quantum mechanics}, Journal of Mathematical Physics \textbf{28}, 1764-1769, 1987.

\bibitem{KS}  S.~Kochen, E.P.~Specker, \emph{The problem of hidden variables in quantum 
mechanics}, Journal of Mathematics and Mechanics \textbf{17}, 59-87, 1967. 

\bibitem{Bell1966} J.S.~Bell, \emph{On the problem of hidden variables in quantum mechanics},
Reviews of Modern Physics \textbf{38}, 447-452, 1966.


\bibitem{BuShill2003} P.~Busch, C.~Shilladay, \emph{Uncertainty reconciles
complementarity with joint measurability}, Physical Review A \textbf{68}, 034102, 1-4, 2003.

\bibitem{BLM1991} P.~Busch, P.J.~Lahti, P.~Mittelstaedt, \emph{The Quantum Theory
of Measurement}, Springer-Verlag, Berlin, 1991, 2nd revised edition 1996.

\bibitem{Busch1987} P.~Busch, \emph{Some realizable joint measurements of
complementary observables}, Foundations of Physics \textbf{17}, 905-937, 1987.


\bibitem{deMuynck-book} W.M.~deMuynck, \emph{Foundations of Quantum Mechanics:
an Empiricist Approach}, Kluwer Academic Publishers, Dordrecht, 2002.


\bibitem{EnglertBergou2000} B-G.~Englert, J.~A.~Bergou, \emph{Quantitative quantum
erasure}, Optics Communications \textbf{179}, 337-355, 2000.

\bibitem{Partha} M.~Krishna, K.R.~Parthasarathy, \emph{An Entropic Uncertainty Principle
for Quantum Measurements}, arXiv:quant-ph/0110025.

\bibitem{MaassenUffink1988} H.~Maassen, J.B.M.~Uffink, \emph{Generalized Entropic
Uncertainty Relations}, Physical Review Letters \textbf{60}, 1103-1106, 1988.

\bibitem{WZ} W.K.~Wootters, W.H.~Zurek, 
\emph{Complementarity in the double-slit experiment: Quantum nonseparability and a quantitative
statement of Bohr's principle},
Physical Review D \textbf{19}, 473-484, 1979.

\bibitem{MPS} P.~Mittelstaedt, A.~Prieur, R.~Schieder, 
\emph{Unsharp particle-wave duality in a photon split beam experiment},
Foundations of Physics \textbf{17}, 891-903, 1987.

\bibitem{GY} D.M.~Greenberger, A.~Yasin, 
\emph{Simultaneous wave and particle knowledge in a neutron interferometer},
Physics Letters A \textbf{128}, 391-394, 1988.


\bibitem{JHS} G.~Jaeger, M.A.~Horne, A.~Shimony, 
\emph{Complementarity of one-particle and two-particle interference},
Physical Review A \textbf{48}, 1023-1027, 1993.

\bibitem{JS} G.~Jaeger, A~.Shimony, 
\emph{Optimal distinction between two non-prthogonal quantum states},
Physics Letters A \textbf{197}, 83-87, 1995.


\bibitem{JSV}  G.~Jaeger, A.~Shimony, L.~Vaidman, 
\emph{Two interferometric complementarities}, 
Physical Review A \textbf{51}, 54-67, 1995.


\bibitem{Carmeli} C.~Carmeli, G.~Cassinelli, E.~DeVito, A.~Toigo, B.~Vacchini,
\emph{A complete characterization of phase space measurements},
Journal of Physics A \textbf{37}, 5057-5066, 2004.

\bibitem{deMHen2001} W.M.~deMuynck, A.J.A.~Hendrikx, \emph{Haroche-Ramsey experiment as
a generalized measurement}, Physical Review A \textbf{63}, 042114/1-15, 2001.

\bibitem{Lahti} P.J.~Lahti, \emph{Uncertainty and Complementarity in Axiomatic
Quantum Mechanics}, International Journal of Theoretical Physics \textbf{19}, 789-842, 1980.

\bibitem{LahtiBug1985} P.J.~Lahti, S.~Bugajski, \emph{Fundamental Principles of
Quantum Theory. II. From a Convexity Scheme to the DHB Theory},
International Journal of Theoretical Physics \textbf{24}, 1051-1080, 1985.

\bibitem{Gudder1973} S.P.~Gudder, \emph{Quantum logics, physical space, position observables
and symmetry}, Reports on Mathematical Physics \textbf{4} 193-202, 1973.



\end{thebibliography}
\end{document}